\documentclass[twocolumn,pra,aps,amsmath,amssymb,showpacs,longbibliography,reprint,floatfix]{revtex4-1}
\usepackage[usenames,dvipsnames]{color}
\usepackage{upgreek,xfrac}
\newcommand{\bigO}{\mathrm{O}}
\usepackage{txfonts}
\usepackage{units}
\usepackage{amsfonts}
\usepackage{amsmath}
\usepackage{amssymb}
\usepackage{mathbbol}
\usepackage{graphicx}
\usepackage{times}
\usepackage{braket}
\usepackage{multirow}
\usepackage{ulem}
\begin{document}

\title{Quantum-ionic features in the absorption spectra of homonuclear diatomic molecules}

\author{A. Crawford-Uranga}
\email{alisonc1986@gmail.com}
\author{D. J. Mowbray}
\email{duncan.mowbray@gmail.com}
\author{D. M. Cardamone}
\email{david.cardamone@gmail.com}
\affiliation{Nano-Bio Spectroscopy group and ETSF Scientific Development Center, Departamento de F\'isica de Materiales, Centro de F\'isica de Materiales CSIC$-$MPC and DIPC, Universidad del Pa\'is Vasco UPV$/$EHU, Avenida de Tolosa 72, E$-$20018, San Sebasti\'an, Spain}
\date{\today}

\begin{abstract}

We show that additional features can emerge in the
linear absorption spectra of homonuclear diatomic molecules when the
ions are described quantum mechanically. 
In particular, the widths and energies of the peaks in the optical spectra change 
with the initial configuration, mass, and charge of the molecule.
We introduce a model that can describe these 
features and we provide a quantitative analysis of the resulting peak energy shifts and width
broadenings as a function of the mass. 

\end{abstract}


\maketitle

\section{Introduction}\label{intro:sec}

Molecular spectroscopy deals with the response of a molecule interacting with an external electromagnetic field. The development of attosecond sources \cite{PaulScience2001,HentschelNature2001,CorkumNatPhys2007} allows one to probe in real time coupled electron-ion dynamics after photoionization processes. Two types of processes are seen in these experiments. The motion of the ions is associated with chemical transformations such as dissociation \cite{KelkensbergPRL2009} in the femtosecond domain. The motion of the electrons is associated with electronic rearrangement processes such as charge redistribution \cite{DrescherNature2002,GouliemakisNature2010}, localization \cite{KlingScience2006,SansoneNature2010} as well as ionization processes such as tunneling \cite{UiberackerNature2007} in the attosecond domain.

Modeling coupled electronic-ionic dynamics in photoionization processes is a formidable challenge for most systems.  For this reason, previous studies have been limited to one (H$_2^+$) and two (H$_2$) electron benchmark systems \cite{SansoneNature2010, QMI_H2+_H2_1D_CONFIGURATION, GrossPRL2001, BOA_SEPARATE, WalshPRA1998}. A full coupled electronic-ionic 3D treatment has only been achieved for the one electron system H$_2^+$, where the ionic motion is confined to the direction of the laser's polarization \cite{QMI_H2+_H2_1D_CONFIGURATION, ChelkowskiPRA1995}. For a full quantum mechanical treatment of two electron two ion systems (H$_2$), it is necessary to confine both the electronic and ionic motion to the laser's polarization direction.  This is a reasonable semiclassical approximation, as the electronic and ionic motion should be predominantly along this direction \cite{QMI_H2+_H2_1D_CONFIGURATION}. Therefore, for most molecules, any quantum-ionic features are typically neglected by instead using classical approximations, e.g., the Born Oppenheimer approximation (BOA) and Ehrenfest dynamics (ED). These approaches rely on a weak coupling between the electronic and ionic wave functions. However, the validity of such approximations breaks down for light atoms, when hybridization between the electronic and ionic wave functions must be included. A quantum versus classical treatment of the ions has been previously used to investigate the localization \cite{SansoneNature2010}, nonsequential double ionization \cite{QMI_H2+_H2_1D_CONFIGURATION}  and harmonic generation \cite{GrossPRL2001} of H$_2$,  as well as the dissociation \cite{BOA_SEPARATE} and proton kinetic energies \cite{WalshPRA1998} for H$_2^+$.

The aim of this paper is a comparison between a quantum mechanical (QMI) and classical (BOA/ED) treatment of the ionic motion to describe coupled electronic and ionic processes \cite{MY_THESIS}. In particular, we consider three and four body systems of electrons and ions for which a fully quantum mechanical treatment of the coupled electron-ion system is feasible. This comparison with respect to the QMI solution is performed both for the static spectra and for the time dependent linear response spectra. In fact, we find significant differences between the QMI and BOA/ED spectra. These features can be quantitatively analyzed using a simple two-level two-parameter model based on the BOA electronic energy levels and the electron-ion mass ratio. The results of our work will help us to determine the domain of applicability of the simplified BOA and ED approaches to interpret coupled electron-ion experiments for more complicated systems.

The paper is organized as follows: in Sec.~\ref{theory:sec}, we introduce the theoretical methods and models employed to simulate the coupled electronic and ionic processes; in Sec.~\ref{methodology:sec}, we explain the methodology to obtain both the ground state and time dependent linear response spectra, as well as the computational details of our calculations; in Sec.~\ref{results:sec}, we show our results for both, the H$_2^+$ and H$_2$ molecules, which we then analyze according to the model we provide; and finally, in Sec.~\ref{conclusions:sec} we summarize the main conclusions and relevant results of our work. Atomic units a.u. ($\hbar = m_e = e = a_0 = 1$) are used throughout, unless stated otherwise.

\section{Theoretical Background}\label{theory:sec}

\subsection{Quantum electron-ion approach}\label{many_body_problem:sec}

A many-body system composed of $N$ ions and $n$ electrons, where both the electrons and ions are treated quantum mechanically (QMI), is described by the total electron-ion time-dependent Hamiltonian
\begin{equation}\label{MB_hamiltonian:eqn}
\hat H(t) = \hat T_I + \hat T_e + \hat V_{\textit{II}} + \hat V_{\textit{Ie}} + \hat V_{\textit{ee}} + \hat V_{\textit{ext}}(t),
\end{equation}
where $\hat T_I$ and $\hat T_e$ are the ionic and electronic kinetic energy operators, respectively, and $\hat V_{\textit{II}}$, $\hat V_{\textit{Ie}}$, $\hat V_{\textit{ee}}$, and $\hat V_{\textit{ext}}(t)$ are the ion-ion, ion-electron, electron-electron, and external potential energy operators, respectively.  The kinetic energy operators take the form
\begin{equation}
\hat T_I = \sum_{\alpha = 1}^N - \frac{1}{2M_{\alpha}} {\bf \nabla}^2_{\alpha},
\end{equation}
where $M_{\alpha}$ is the mass of ion $\alpha$, and
\begin{equation}\label{kinetic_electron:eqn}
\hat T_e = \sum_{i = 1}^n - \frac{1}{2} {\bf \nabla}^2_{i}.
\end{equation}
The interaction between the ions is given by
\begin{equation}\label{ion_ion:eqn}
\hat V_{\textit{II}} = \frac{1}{2} \sum_{\substack{\alpha,\beta = 1 \\ \alpha \neq \beta}}^N \frac{Q_{\alpha} Q_{\beta}}{|{\bf R}_{\alpha}-{\bf R}_{\beta}|},
\end{equation}
where $Q_{\alpha}$, $Q_{\beta}$, {\bf R}$_{\alpha}$ and {\bf R}$_{\beta}$ are the corresponding charges and positions of ion $\alpha$ and $\beta$. Similarly, the electron-electron repulsion is
\begin{equation}\label{electron_electron:eqn}
\hat V_{\textit{ee}} = \frac{1}{2} \sum_{\substack {i,j = 1 \\ i \neq j}}^n \frac{1}{|{\bf r}_{i}-{\bf r}_{j}|},
\end{equation}
where {\bf r}$_i$ and {\bf r}$_j$ are the positions of electrons $i$ and $j$, while the interaction between electrons and ions is
\begin{equation}\label{ion_electron:eqn}
\hat V_{\textit{Ie}} = - \sum_{\alpha = 1}^N \sum_{i=1}^{n}  \frac{Q_{\alpha}}{|{\bf r}_{i}-{\bf R}_{\alpha}|}.
\end{equation}
Finally, $\hat V_{\textit{ext}}(t)$ describes the interaction of the system of electrons and ions with an external electromagnetic time-dependent field, defined explicitly in Sec.~\ref{spectra:sec}.

The time-dependent Schr\"{o}dinger equation takes the form
\begin{equation}\label{MB_depend:eqn}
\begin{split}
i \frac{\partial\psi}{\partial t} &=
i \frac{\partial}{\partial t} \psi({\bf R}_1 S_1,{\bf R}_2 S_2,...{\bf R}_N S_N;{\bf r}_1 s_1,{\bf r}_2 s_2...{\bf r}_n s_n,t)\\ &=
 \hat H(t) \psi({\bf R}_1 S_1,{\bf R}_2 S_2,...{\bf R}_N S_N;{\bf r}_1 s_1,{\bf r}_2 s_2...{\bf r}_n s_n,t),
\end{split}
\end{equation}
where $\psi$ is the time-dependent electron-ion wavefunction. This depends on the positions ${\bf R}_{\alpha}$ and ${\bf r}_i$ and on the spin coordinates $S_{\alpha}$ and $s_{i}$ of ion $\alpha$ and electron $i$, respectively. 

For time-independent problems ($\hat V_{\textit{ext}}(t)=0$), the general solution of the time-dependent Schr\"{o}dinger equation can be written as 
\begin{equation}\label{TISE:eqn}
\psi = \sum_k c_k e^{-i\varepsilon_k t} 
\psi _k({\bf R}_1 S_1,{\bf R}_2 S_2,...{\bf R}_N S_N;{\bf r}_1 s_1,{\bf r}_2 s_2...{\bf r}_n s_n)
\end{equation}
where $\varepsilon_k$ and $\psi_k$ are the $k^{th}$ eigenvalue and eigenstate of the electron-ion stationary Schr\"{o}dinger equation 
\begin{equation}\label{MB_indep:eqn}
\begin{split}
\hat H \psi_k & = \hat H \psi_k({\bf R}_1 S_1,{\bf R}_2 S_2,...{\bf R}_N S_N;{\bf r}_1 s_1,{\bf r}_2 s_2...{\bf r}_n s_n)\\ & = 
\varepsilon_k \psi_k({\bf R}_1 S_1,{\bf R}_2 S_2,...{\bf R}_N S_N;{\bf r}_1 s_1,{\bf r}_2 s_2...{\bf r}_n s_n),
\end{split}
\end{equation}
with
\begin{equation}\label{USE:eqn}
\hat H = \hat T_I + \hat T_e + \hat V_{\textit{II}} + \hat V_{\textit{Ie}} + \hat V_{\textit{ee}}.
\end{equation}
We will next focus on the time-independent solution until introducing an external field in Sec.~\ref{spectra:sec}.

Solving the QMI problem is very demanding computationally for many-body systems.  In fact, it quickly becomes unfeasible for systems with more than three independent variables.  For this reason, we restrict consideration herein to one or two-electron diatomic molecules whose motion is confined to one direction (see Sec.~\ref{model_1D:sec}).  By applying an appropriate coordinate transformation, such systems may be modeled with only two or three independent variables (see Appendix~\ref{AppendixA}).  In Secs.~\ref{BOA:sec} and \ref{ED:sec}, we introduce two of the most widely used approximations to simplify the general many-body electron-ion problem. 

\subsection{Born-Oppenheimer approximation}\label{BOA:sec}

Within the Born-Oppenheimer approximation (BOA) \cite{BOA_ORIGINAL}, the total electronic-ionic wavefunction $\psi$ is assumed to be separable into an ionic $\chi$ and electronic $\varphi$ part. As the electrons move much faster than the ions, we assume that the kinetic energy of the ions does not cause the excitation of the electrons to another electronic state, i.e., an adiabatic approximation.   Such an approximation is valid so long as the ratio of vibrational to electronic energies, $E_{\textit{vib}}$ to $E_{\textit{elec}}$, which goes as the root of the electron-ion mass ratio, i.e., $ E_{\textit{vib}}/E_{\textit{elec}} \approx \sqrt{m_e/M}$, is small \cite{BOA_ORIGINAL}(see Appendix~\ref{AppendixB} for details).  Since for a proton $M_{\mathrm{p}} \approx \mathrm{1836}m_e$ and  $E_{\textit{vib}}/E_{\textit{elec}} \sim 0.02$, the BOA is expected to work quite well for our molecules. We thus may neglect $\hat T_I = 0$ from Eq.~(\ref{USE:eqn}), although the electrons still feel  the static field of the ions ($\hat V_{eI}, \hat V_{\textit{II}} \neq 0$).

The separable BOA solution $\psi$ of the electron-ion stationary Schr\"{o}dinger equation (\ref{TISE:eqn}) is given by \cite{BOA_SEPARATE}
\begin{equation}\label{separate:eqn}
\psi  = \chi ({\bf R}_1 S_1,{\bf R}_2 S_2,...{\bf R}_N S_N) 
\varphi^{({\bf R}_1,{\bf R}_2,...{\bf R}_N)}({\bf r}_1 s_1,{\bf r}_2 s_2...{\bf r}_n s_n),
\end{equation}
where $\chi$ depends on the ionic coordinates only and $\varphi$ depends on both the electronic coordinates and on the ionic coordinates which, however, only enter into the electronic wavefunctions as parameters. As shown in Ref. \cite{BOA_SEPARATE}, this may be done for the Hamiltonian of Eq.~\ref{MB_hamiltonian:eqn} without loss of generality.

If we insert Eq.~(\ref{separate:eqn}) directly into Eq.~(\ref{MB_indep:eqn}), we obtain the general coupled electron-ion BOA problem
\begin{equation}\label{BOA_exp:eqn}
\begin{split}
\hat H \psi_{k}  =&  -  \sum_{\alpha = 1}^N  \frac{{\nabla}^2_{\alpha} \chi}{2M_{\alpha}}   \varphi - \chi \sum_{\alpha = 1}^N  \frac{{\nabla}^2_{\alpha} \varphi}{2M_{\alpha}}  - \sum_{\alpha = 1}^N  \frac{{\bf \nabla}_{\alpha} \chi \cdot {\bf \nabla}_{\alpha} \varphi}{M_{\alpha}}  \\
& + \chi \left(\sum_{i = 1}^n - \frac{\nabla_{i}^2}{2} + \hat V_{\textit{Ie}} + \hat V_{\textit{ee}} + \hat V_{\textit{II}}\right) \varphi \\
=&\ E_{k} \psi_{k}.
\end{split}
\end{equation}
However, one normally separates Eq.~(\ref{BOA_exp:eqn}) into an electronic problem only in $\varphi$ and an ionic problem only in $\chi$.  To do so, one first solves the electronic-only BOA frozen ion Schr\"{o}dinger equation, where the ionic coordinates ${\bf R}_{\alpha}$ only enter as fixed parameters in $\varphi$:
\begin{equation}\label{general_He:eqn}
\begin{split}
\hat H_e \varphi_i &= \hat H_e \varphi_i^{({\bf R}_1,{\bf R}_2,...{\bf R}_N)}({\bf r}_1 s_1,{\bf r}_2 s_2...{\bf r}_n s_n)\\
& = 
E_i ({\bf R}_1,{\bf R}_2,...{\bf R}_N) \varphi_i^{({\bf R}_1,{\bf R}_2,...{\bf R}_N)}({\bf r}_1 s_1,{\bf r}_2 s_2...{\bf r}_n s_n),
\end{split}
\end{equation}
where
\begin{equation}
\hat H_e = \hat T_e + \hat V_{\textit{Ie}} + \hat V_{\textit{ee}} + \hat V_{\textit{II}}.
\end{equation}
In this way, one may find the so-called $i^{th}$ potential energy surfaces $E_i ({\bf R}_1,{\bf R}_2,...{\bf R}_N)$ (PES).  These are representations of the electronic energy landscape as a function of the ionic coordinates.

In the next step, the ionic BOA Schr\"{o}dinger equation is solved by adding the previously neglected kinetic energy of the ions to the potential energy surfaces obtained from the frozen ion  Schr\"{o}dinger equation
\begin{equation}\label{ionic_general:eqn}
\hat H_I^i \chi_{ij}({\bf R}_1 S_1,{\bf R}_2 S_2,...{\bf R}_N S_N) \\
= E_{ij} \chi_{ij}({\bf R}_1 S_1,{\bf R}_2 S_2,...{\bf R}_N S_N),
\end{equation}
where
\begin{equation}\label{ionic_value:eqn}
\hat H_I^i = \sum_{\alpha=1}^{N} - \frac{1}{2M_{\alpha}}{\nabla^2_{\alpha}} + E_i({\bf R}_1,{\bf R}_2,...{\bf R}_N),
\end{equation}
and the ionic excitations $j$ depend on the electronic excitations, $i$.

Comparing Eq.~(\ref{BOA_exp:eqn}) with Eq.~(\ref{ionic_value:eqn}) we realize that the second and third terms of Eq.~(\ref{BOA_exp:eqn}) are neglected in the BOA.  This is because we assume that the kinetic energy of the ions is not affecting the electronic part $\varphi$, i.e., ${\bf \nabla}_\alpha\varphi \approx 0$.

\subsection{Ehrenfest dynamics}\label{ED:sec}

Within the Ehrenfest dynamics (ED) scheme \cite{ED_ORIGINAL}, we solve the coupled evolution of the electrons and ions. The electrons evolve quantum mechanically, whereas the ions evolve classically on a mean time-dependent PES $\varphi_i(t)$ weighted by the different BOA PES $\varphi_i$ in Eq.~(\ref{general_He:eqn})
\begin{equation}\label{BOA_ED:eqn}
\varphi_i({\bf R}_{\alpha}(t)) = \sum_{i}^{n} c_i(t) \varphi_i
\end{equation}
 
The ions are evolved according to Newton's equation of motion
\begin{equation}
  {\bf F}_{\textit{ED}}({\bf R}_{\alpha}(t)) = M_{\alpha}\frac{d^2 \mathbf{R}_{\alpha}(t)}{dt^2}
\end{equation}
which satisfies the following potential energy derivative condition
\begin{equation}\label{classicalforces:eqn}
\begin{split}
{\bf F}_{\textit{ED}}({\bf R}_{\alpha}(t)) &= - \sum_{i}^{n} \left|c_i(t)\right|^2 \vec{\nabla}_\alpha \varepsilon_i \left({\bf R}_{\alpha}(t)\right) \\
&= - \left \langle \varphi (t) \left | \vec{\nabla}_\alpha H_{e}({\bf R}_{\alpha}(t)) \right | \varphi (t) \right \rangle
\end{split}
\end{equation}
where Eq.~(\ref{BOA_ED:eqn}) and the Hellmann-Feynman theorem have been employed.  The Ehrenfest electron-ion scheme consists of the time propagation of the coupled Eqs.~(\ref{general_He:eqn}) and (\ref{classicalforces:eqn}).

\subsection{Model systems: Initial configurations and Hamiltonians}\label{model_1D:sec}

We model the positively charged one electron H$_2^+$ and neutral two electron H$_2$ homonuclear diatomic molecules, assuming their motion is confined to one direction.  Such a model should provide a reasonable description of a molecule excited by a laser field, where the electronic and ionic motion are confined to the polarization axis of the laser field \cite{BOA_SEPARATE}.  In this case the QMI problem described in Sec.~\ref{many_body_problem:sec}, where both electrons and ions are treated quantum mechanically, can be solved exactly.  Furthermore, by working in center of mass coordinates, the computational effort required to solve Eq.~(\ref{MB_depend:eqn}) is significantly reduced. 

However, the singularity in the bare Coulomb interaction of Eqs.~(\ref{ion_ion:eqn}), (\ref{electron_electron:eqn}), and (\ref{ion_electron:eqn}) in 1D makes the direct numerical solution of the Schr\"{o}dinger equation (\ref{MB_depend:eqn}) unfeasible.  Instead, one employs the so-called ``soft Coulomb interaction'' \cite{SOFT_COULOMB,BOA_SEPARATE}.  For two particles $i$ and $j$ with charges $Q_i$ and $Q_j$, the soft Coulomb interaction $V_{int}$ has the general form
\begin{equation}\label{soft_coulomb:eqn}
V_{int}(s) = \frac{Q_i Q_j}{\sqrt{s^2 + \Delta^2}},
\end{equation}
where $s$ is the separation between the two charges and $\Delta$ is the soft Coulomb parameter~\cite{SOFT_COULOMB}.  Typically, $\Delta = a_0$, although other values can also be used \cite{QMI_H2+_H2_1D_CONFIGURATION}.

\begin{figure}%[!t]
\begin{center}
\includegraphics[scale=0.3]{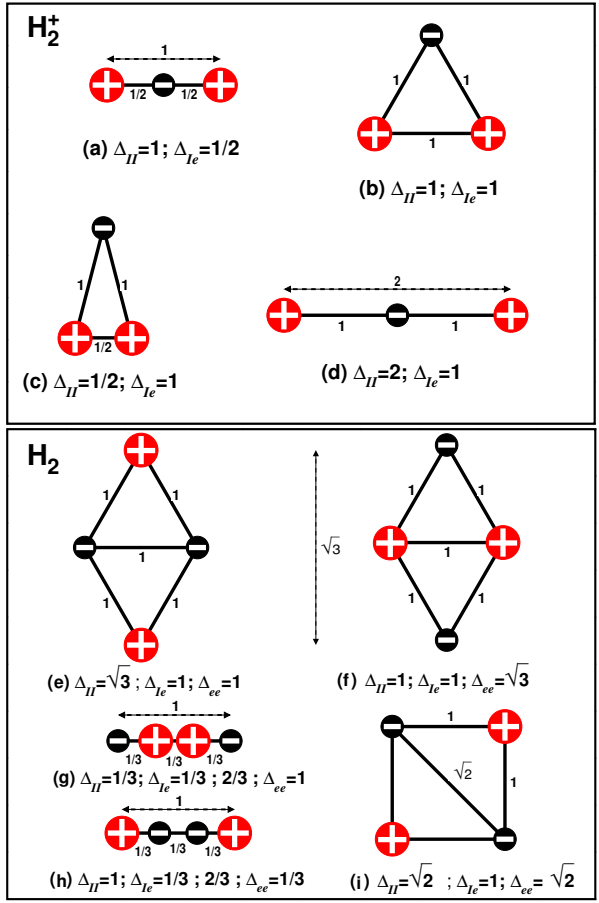}
\end{center}
\caption{\footnotesize
(Color online) Schematic representation of the (a--d) H$_2^+$ and (e--i) H$_2$ geometries for the minimum ionic separation $\Delta_{\textit{II}}$, ion--electron separation $\Delta_{\textit{Ie}}$, and electron--electron separation $\Delta_{\textit{ee}}$ for each configuration. Protons are shown in red and electrons in black.}
\label{EPS_VALUES:fig}
\end{figure}

In essence, the soft Coulomb interaction amounts to a displacement of the trajectories of the two particles in an orthogonal direction.  So for a hydrogen atom, a soft Coulomb interaction of 
\begin{equation}
V_{int}(s) = - \frac{1}{\sqrt{s^2 + a_0^2}},
\end{equation}
is equivalent to having a bare Coulomb interaction with the electron and proton trajectories required to be parallel, with a minimum separation of $a_0$.  This is a quite reasonable assumption, as the most probable electron-proton separation in a hydrogen atom is the Bohr radius $a_0$.  Soft Coulomb parameters correspond to the separation between the 1D trajectories that each electron and ion will move along in 3D with a bare Coulomb interaction. We may directly map the 1D soft Coulomb problem to a bare Coulomb problem in 3D where the electrons and ions are separated by the soft Coulomb parameter distances shown in Fig.~\ref{EPS_VALUES:fig}, while their motion is confined in one direction.  In this way, one clearly sees that constraint rotations of the molecules are possible in 3D, while their motion is still confined to 1D. As discussed in Sec.~\ref{spectra:sec}, the molecules will be perturbed by a kick confined in one direction. 

In Fig.~\ref{EPS_VALUES:fig} we show the various configurations we have employed to model an H$_2^+$ or H$_2$ molecule whose electronic and ionic motion is confined to one direction.  These configurations are specified by the soft Coulomb parameters between the ions $\Delta_{\textit{II}}$, the electrons $\Delta_{\textit{ee}}$ and the ions and electrons $\Delta_{\textit{Ie}}$. One such configuration has been used previously \cite{QMI_H2+_H2_1D_CONFIGURATION} to study the dynamics of a one-dimensional H$_2$ model molecule in strong laser fields by means of QMI.

An analysis of the effect of the initial configuration on the optical spectra is shown in Sec.~\ref{results:sec}.  The classical energies of positively charged and neutral homonuclear diatomic molecules whose motion is confined to one direction are given by
\begin{equation}\label{E_ini:eqn}
\begin{split}
E =&\ \frac{1}{2} M V_{1}^{2}  +  \frac{1}{2} M V_{2}^{2}  +  \frac{1}{2} v_1^2
  - \frac{1}{\sqrt{(x_1-X_1)^2+\Delta_{\textit{Ie}}^2}} \\
  &-
   \frac{1}{\sqrt {(X_2-x_1)^2 +\Delta_{\textit{Ie}}^2}} 
  + \frac{1}{\sqrt {(X_2 - X_1)^2+\Delta_{\textit{II}}^2}}\;,
\end{split}
\end{equation}
and
\begin{equation}\label{E_ini_hydrogen:eqn}
\begin{split}
E =&\ \frac{1}{2} M V_{1}^{2}  +  \frac{1}{2} M V_{2}^{2}  +  \frac{1}{2} v_{1}^{2}  +  \frac{1}{2} v_{2}^{2}\\& -
    \frac{1}{\sqrt{(x_1-X_1)^2+\Delta_{\textit{Ie}}^2}} - \frac{1}{\sqrt{(X_2-x_2)^2+\Delta_{\textit{Ie}}^2}} 
   \\&- \frac{1}{\sqrt{(x_2-X_1)^2+\Delta_{\textit{Ie}}^2}} - \frac{1}{\sqrt {(X_2-x_1)^2 +\Delta_{\textit{Ie}}^2}}\\& +
    \frac{1}{\sqrt {(X_2 - X_1)^2+\Delta_{\textit{II}}^2}} + \frac{1}{\sqrt {(x_2 - x_1)^2+\Delta_{\textit{ee}}^2}}, 
\end{split}
\end{equation}
respectively.  Here, $M$ is the ion mass; $V_1$, $V_2$, $X_1$ and $X_2$ are the ionic velocities and positions for both molecules along the direction of motion along the direction of motion.

The first three and four terms of Eqs.~(\ref{E_ini:eqn}) and (\ref{E_ini_hydrogen:eqn}) are the kinetic energies of the electrons and ions in the molecules, as explained above. The remaining terms correspond to the attractive and repulsive electrostatic potential energy terms between such electrons and ions.

The spatial configuration of positively charged or neutral homogeneous diatomic molecules in 1D does not change if the particle positions are translated uniformly. This reduces our three- and four-body coordinate problems into two- and three-body ones, respectively.

We rewrite the classical energies in Eqs.~(\ref{E_ini:eqn}) and (\ref{E_ini_hydrogen:eqn}) in terms of the center-of-mass transformation \cite{CM} (see Appendix \ref{AppendixA}) to obtain the  following two-body ($X$,$\xi$) and three-body ($X$,$x$,$\xi$) Hamiltonians 
\begin{equation}\label{quantum_H:eqn}
\begin{split}
\hat H(X,\xi) =& - \frac{1}{M} \frac{\partial^2}{\partial{X}^2} -  
  \frac{2M+1}{4M} \frac{\partial^2}{\partial{\xi}^2}-
  \frac{1}{\sqrt{(\frac{X}{2} +\xi)^2 + \Delta_{\textit{Ie}}^2}}  
  \\ &-\frac{1}{\sqrt{(\frac{X}{2} -\xi)^2 + \Delta_{\textit{Ie}}^2}} + 
  \frac{1}{\sqrt{X^2 + \Delta_{\textit{II}}^2}},
\end{split}
\end{equation}
and
\begin{equation}\label{H_int:eqn}
\begin{split}
\hat H(X,x,\xi) =&
  - \frac{1}{M} \frac{\partial^2}{\partial{X}^2} -  
  \frac{\partial^2}{\partial{x}^2} -  
  \frac{1+M}{4M} \frac{\partial^2}{\partial{\xi}^2} \\&-
  \frac{1}{\sqrt{(\frac{X}{2} - \frac{x}{2} + \xi)^2 + \Delta_{\textit{Ie}}^2}}
  - \frac{1}{\sqrt{(\frac{X}{2} - \frac{x}{2} - \xi)^2 + \Delta_{\textit{Ie}}^2}} \\
   &- \frac{1}{\sqrt{(\frac{X}{2} + \frac{x}{2} + \xi)^2 + \Delta_{\textit{Ie}}^2}} 
  - \frac{1}{\sqrt{(\frac{X}{2} + \frac{x}{2} - \xi)^2 + \Delta_{\textit{Ie}}^2}} \\&+ 
   \frac{1}{\sqrt{x^2 + \Delta_{\textit{ee}}^2}}
  + \frac{1}{\sqrt{X^2 + \Delta_{\textit{II}}^2}},
\end{split}
\end{equation}
for positively charged and neutral homogeneous diatomic molecules, respectively, after removing the center of mass term. Here $X$ and $x$ are the ionic and electronic separations and $\xi$ is the separation between ionic and electronic centers of mass along the direction in which their motion is confined.

Although the electrons are treated quantum mechanically along their direction of motion, the confinement of their motion and position along one direction is inherently classical.  For this reason, our treatment herein is essentially semiclassical: quantum mechanical along the direction of motion, and classical perpendicular to the direction of motion.  This has important repercussions for the H$_2$ configurations shown in Fig.~\ref{EPS_VALUES:fig}(g) and (h).  For these cases, the Hamiltonian of Eq.~(\ref{H_int:eqn}) is no longer symmetric under ion or electron exchange, since $\Delta_{\textit{Ie}}$ is either $\frac{1}{3}a_0$ or $\frac{2}{3}a_0$.  This reflects the limitations of such a semiclassical treatment.  So although the Hamiltonian of Eq.~(\ref{H_int:eqn}) is still symmetric under ion or electron exchange for the H$_2$ configurations shown in Fig.~\ref{EPS_VALUES:fig}(e), (f), and (i), the confinement of the electron's position perpendicular to its motion may still have an important impact for these configurations.  

Our aim here is to assess the accuracy of the approximations introduced in Secs.~\ref{BOA:sec} and \ref{ED:sec}, in which the ions are treated classically.  To accomplish this, we vary the ionic mass $M$ in our homogeneous diatomic molecules for the many-body problem, while fixing the the ionic charge $Q = e$.   We only consider $Q = e$ because the repulsion between the ions of more massive homonuclear diatomic molecules with a single electron would be so large that the molecules would be unstable \cite{UNSTABLE}.  Furthermore, this allows us to directly compare absorption spectra between these model systems for a fixed interaction potential.

\subsection{Symmetries of the many-body wavefunction}\label{parity:sec}

Since for our positively charged homogeneous diatomic molecules there are one electron and two ions, the antisymmetry of the many-body wavefunction must be enforced for the ions only as
\begin{equation}
\psi (X_1 S_1,X_2 S_2,x s) = - \psi (X_2 S_2,X_1 S_1,x s),
\end{equation}
for the triplet and 
\begin{equation}
\psi (X_1 S_1,X_2 S_2,x s) =  \psi (X_2 S_2,X_1 S_1,x s),
\end{equation}
for the singlet.

For our neutral homogeneous diatomic molecules there are two electrons and two ions. Therefore, the antisymmetry of the many-body wavefunction must be enforced both for the ions and the electrons as
\begin{equation}
\psi (X_1 S_1,X_2 S_2,x_1 s_1,x_2 s_2) = - \psi (X_2 S_2,X_1 S_1,x_1 s_1,x_2 s_2),
\end{equation}
for the ionic triplet,
\begin{equation}
\psi (X_1 S_1,X_2 S_2,x_1 s_1,x_2 s_2) =  \psi (X_2 S_2,X_1 S_1,x_1 s_1,x_2 s_2),
\end{equation}
for the ionic singlet,
\begin{equation}
\psi (X_1 S_1,X_2 S_2,x_1 s_1,x_2 s_2) = - \psi (X_1 S_1,X_2 S_2,x_2 s_2,x_1 s_1),
\end{equation}
for the electronic triplet, and
\begin{equation}
\psi (X_1 S_1,X_2 S_2,x_1 s_1,x_2 s_2) = - \psi (X_1 S_1,X_2 S_2,x_2 s_2,x_1 s_1),
\end{equation}
for the electronic singlet, respectively.

Therefore, due to the exchange symmetry of the many-body wavefunction, in order to have a 
total antisymmetric many-body wavefunction, the spatial part of the ionic and electronic wavefunction must be odd for the triplet and even for the singlet under the exchange of two identical particles.  Consequently, we will only be concerned with the spatial part of the wavefunction, with the spin part already being separated off due to the exchange symmetry of the many-body wavefunction.

\section{Methodology}\label{methodology:sec} 

\subsection{Ground state}\label{BOA_molecules:sec}

The QMI eigenvalues are obtained by inserting Eqs.~(\ref{quantum_H:eqn}) and (\ref{H_int:eqn}) into Eq.~(\ref{MB_indep:eqn}) for the H$_2^+$ and H$_2$ molecules, respectively. 

To obtain the PES within the BOA and ED we would insert Eqs.~(\ref{quantum_H:eqn}) and 
(\ref{H_int:eqn}), neglecting the first term, into Eq.~(\ref{general_He:eqn}) for the H$_2^+$ and H$_2$ molecules, respectively. For the BOA and ED ground state electron-ion level, we do not compute Eq.~(\ref{ionic_general:eqn}). Instead, we fit the ground state PES around its minimum energy at the inter-ionic distance $X_{\textit{eq}}$ using a harmonic approximation $E_{\textit{gs}}(X_{\textit{eq}}) + \frac{1}{2}k_1 {(X-X_{\textit{eq}})}^2$, where $k_1=\omega_I^2 \mu_p$ is the harmonic constant, $\omega_I$ is the harmonic oscillator vibrational frequency and $\mu_p$ is the ionic reduced mass defined in Eq.~(\ref{reduced_mass_electron:eqn}). From $\omega_I$, we obtain the ground state electron-ion eigenvalue of a harmonic oscillator $\varepsilon_{\textit{gs}}^{\textit{BOA/ED}} = E_{\textit{gs}}(X_{\textit{eq}}) + \frac{1}{2}\omega_I$ in the BOA and ED PES picture. 

The inversion symmetry with respect to the inter-ionic $X$ coordinate of the potential in Eqs.~(\ref{quantum_H:eqn}) and (\ref{H_int:eqn}), leads to a doubly-degenerate solution $\varepsilon_k$ for each state $\psi_k$ in Eq.~(\ref{MB_indep:eqn}), for sufficiently bound global ground state potentials. The inversion symmetry with respect to the inter-electronic $x$ coordinate of the potential in Eqs.~(\ref{quantum_H:eqn}) and (\ref{H_int:eqn}) is not related to the statistics of the ions, but to the symmetry of the electronic molecular orbital.

\subsection{Time dependent linear response spectra}\label{spectra:sec}

To obtain the linear response photoabsorption spectra we apply an initial impulsive perturbation, or ``kick'' \cite{KICK}
\begin{equation}\label{dipole_coord_old:eqn}
\begin{split}
\mathcal{K}(\mathrm{H}_2^+) &= e^{i K(X_1 + X_2 -x)}, \\ 
\mathcal{K}(\mathrm{H}_2)   &= e^{i K(X_1 + X_2 - x_1 - x_2)},
\end{split}
\end{equation}
to the ground state wavefunctions $\psi_{\textit{gs}}$ of our H$_2^+$ and H$_2$ molecules, respectively, for the BOA and QMI approaches. $K$ is a measure of the strength of the kick. We employ a converged kick strength of $K=0.001$, for which the linear response spectra does not change if it is decreased further. Using the center of mass coordinates defined in Appendix \ref{AppendixA}, the terms in Eq.~(\ref{dipole_coord_old:eqn}) become 
\begin{equation}\label{dipole_coord_new:eqn}
\begin{split}
\mathcal{K}(\mathrm{H}_2^+) &=e^{i K\left(X_{\mathrm{CM}_2} - \frac{2M+2}{2M+1}\xi\right)}, \\ 
\mathcal{K}(\mathrm{H}_2) &=e^{-i K2\xi},
\end{split}
\end{equation}
where $X_{\mathrm{CM}_2}$ is the global center of mass coordinate, and $\xi$ is the separation between the ionic and electronic centers of mass along the direction in which their motion is confined. The perturbative kick $\mathcal{K}$ will only induce polarization on the coordinates $\xi$ defined for H$_2^+$ and H$_2$ in Eqs.~(\ref{MU:eqn}) and (\ref{JUJU:eqn}) for the BOA and QMI methods.

In linear response, we expand Eq.~(\ref{dipole_coord_new:eqn}) in terms of $K$, neglecting higher order terms
\begin{equation}\label{dipole_coord_final:eqn}
\begin{split}
\mathcal{K}(\mathrm{H}_2^+) &\approx 1 + i K\left(X_{\mathrm{CM}_2} - \frac{2M+2}{2M+1}\xi\right),\\ 
\mathcal{K}(\mathrm{H}_2) &\approx 1 - i K2\xi.
\end{split}
\end{equation}

For the ED approach one should follow the same procedure starting from Eq.~(\ref{dipole_coord_old:eqn}), but substituting the electronic coordinates $x$ for $-\frac{2M+2}{2M+1}\xi \approx -\xi \, \mathrm{when} \, M \gg 1$ for H$_2^+$ and $x_2 + x_1$ for $-2\xi$ for H$_2$. During the time propagation the ions are not kicked, but evolve as parameters according to Eq.~(\ref{classicalforces:eqn}). In this case, the electron is kicked relative to the center of mass of the ions for the H$_2^+$ molecule, and the two electrons are kicked relative to their distance to the ions for the H$_2$ molecule. However, the linear response absorption spectra does not depend on uniform translations of the ions and electrons.

The enforced time-reversal symmetry evolution operator \cite{TRES} we apply to propagate our equations after this external perturbation has been applied is given by 
\begin{equation}
U(t + \Delta t, t) = e^{-i \frac{\Delta t}{2} H(t + \Delta t)} \times e^{-i \frac{\Delta t}{2} H(t)},
\end{equation}
where the Hamiltonian $H(t+\Delta t)$ is calculated from
\begin{equation}\label{33}
\psi(t+\Delta t) = e^{-i\Delta t H(t)} \psi(t), 
\end{equation}
and the kicked initial state we propagate is
\begin{equation}\label{EQ_TERM:eqn}
\psi(\Delta t) = e^{-i \Delta t H_0} \mathcal{K} \psi_{\textit{gs}},
\end{equation}
where $\psi_{\textit{gs}}$ is the ground state eigenstate of the time independent Hamiltonian $H_0$ of Eqs. (\ref{quantum_H:eqn}) and (\ref{H_int:eqn}) for H$^+_2$ and H$_2$ , respectively.

The time dependent Hamiltonian $H(t)$ is then obtained by time propagation at each time step self consistently according to Eq.~(\ref{33}), starting from the kicked initial state given in Eq.~(\ref{EQ_TERM:eqn}). The expectation value of the dipole moment $d$ at time $t$ is
\begin{equation}\label{dt:eqn}
d(t) = \langle \psi(t)| \hat \xi | \psi(t) \rangle.
\end{equation}
If we assume that the Hamiltonian does not evolve in time and we insert Eq. (\ref{dipole_coord_final:eqn}) into Eq. (\ref{EQ_TERM:eqn}), using the completeness relation $\sum_k |\psi_k \rangle \langle \psi_k | = \mathbb{1}$ and Eq. (\ref{TISE:eqn}) we get
\begin{equation}\label{TD_pw:eqn}
|\psi(t)\rangle \approx  e ^{-i\varepsilon_{\textit{gs}} t} |\psi_{\textit{gs}}\rangle - 
i K \sum_k e ^{-i\varepsilon_{k} t} \left \langle \psi_k \left | \frac{2M+2}{2M+1}\hat{\xi}  \right| \psi_{\textit{gs}} 
\right \rangle | \psi_k   \rangle
\end{equation}
for the H$_2^+$ molecule and 
\begin{equation}\label{TD_pw_H2:eqn}
\begin{split}
 |\psi(t) \rangle &\approx  e ^{-i\varepsilon_{\textit{gs}} t} |\psi_{\textit{gs}} \rangle  - i K \sum_k e ^{-i\varepsilon_{k} t} 
 \left \langle \psi_k \left| 2\hat{\xi}  \right| \psi_{\textit{gs}} \right \rangle | \psi_k \rangle.
\end{split}
\end{equation}
for the H$_2$ molecule. From Eqs.~(\ref{TD_pw:eqn}) and (\ref{TD_pw_H2:eqn}), we see that only the $\psi_{\textit{gs}}$ to odd $k$ dipole moment matrix elements are non-zero by symmetry, i.e parity, since $\hat \xi$ is an odd operator.

Eq.~(\ref{dt:eqn}) can be written as
\begin{equation}\label{euler:eqn}
d(t)\ \approx -2K \frac{2M+2}{2M+1} \sum_k \sin \omega_k t \left| \left \langle \psi_{k} \left | \hat \xi \right | \psi_{\textit{gs}} \right \rangle \right|^2,
\end{equation}
for the H$_2^+$ molecules using Eq.~(\ref{TD_pw:eqn}) and 
\begin{equation}\label{euler2:eqn}
d(t)\ \approx -4K 
\sum_k \sin\omega_k t \left| \left \langle \psi_{k} \left | \hat \xi \right | \psi_{\textit{gs}} \right \rangle \right |^2\!,
\end{equation}
for the H$_2$ molecules using Eq.~(\ref{TD_pw_H2:eqn}), where $d(t)$ depends linearly on $K$ and $\omega_k \equiv \varepsilon_k - \varepsilon_{\textit{gs}}$.  

However, in the ED approach the ionic coordinates are updated at each time step.  This makes $\hat{\xi}$ and the Hamiltonian time dependent.  For this reason, the dipole moment from the ED approach does not necessarily have the form of Eqs.~(\ref{euler:eqn}) and (\ref{euler2:eqn}).  As we will show in Sec. \ref{H2results:sec}, the time-dependent effects of zero-point motion within ED, which are incorporated into the coordinate $\xi(t) = x - X_{\textrm{CM}_{1}}$, have an important impact on the spectra.

The optical photoabsorption cross section spectra $\sigma_{abs}$ is obtained by performing a discrete Fourier transform of $d(t)$ \cite{FT}.  More precisely, 
\begin{equation}\label{FT:eqn}
\sigma_{abs} = 4 \pi \alpha \omega {\rm Im}\left[ \frac{1}{K} \sum_{t=0}^T \Delta t e^{-i \omega t} f\left(\frac{t}{T}\right) \left[d(t)-d(0)\right]\right],
\end{equation}
where 
\begin{equation}\label{THIRD_ORDER_POL:eqn}
f(x) = e^{-25x^2},
\end{equation}
is a Gaussian damping applied to improve the resolution of the photoabsorption peaks, $\omega$ is the frequency of the oscillations of $d(t)$, $\alpha$ is the fine structure constant, $T=1000$ is the total propagation time, and $\Delta t=0.01$ is the time step.

\subsection{Computational details}

All numerical calculations have been performed using the real space electronic structure code \texttt{Octopus} \cite{OCTOPUS}. We discretize the configuration space of the H$_2^+$ and H$_2$ molecules, using a finite set of values (i.e. a so-called grid) for the coordinates $X$, $x$ and $\xi$ in the box intervals $X \in [-L_X,L_X]$, $x \in [-L_x,L_x]$ and $\xi \in [-L_\xi,L_\xi]$.
These are discretized as 
\begin{equation}
\begin{split}
X_i &= -L_X + i\Delta X \ {\mbox {for}} \ i=0,1,2...N_X,\\
x_j &= -L_x + j\Delta x \ {\mbox {for}} \ j=0,1,2...N_x,\\
\xi_k &= -L_\xi + k\Delta \xi \ {\mbox {for}} \ k=0,1,2...N_\xi,
\end{split}
\end{equation}
using $N_X$, $N_x$ and $N_{\xi}$ equally spaced points, respectively. The spacing between two adjacent points in the $X$, $x$ and $\xi$ directions are $\Delta X = \frac{\mathrm{2}L_X}{N_X}$, $\Delta x = \frac{\mathrm{2}L_x}{N_x}$, $\Delta\xi = \frac{\mathrm{2}L_\xi}{N_\xi}$.  Convergence is achieved when a decrease in $\Delta X$, $\Delta x$, $\Delta \xi$ and an increase in $L_X$, $L_x$, $L_{\xi}$ does not change the electron-ion static and time propagation linear response spectra.

For the H$_2^+$ type molecules, ground state convergence is achieved for $L_X=L_{\xi}=10a_0$, $\Delta X=0.05a_0$ and $\Delta \xi =0.1a_0$. To obtain the PES we have used $L_\xi=100a_0$ and $\Delta \xi=0.1a_0$.   Generally, the convergence of the QMI optical spectra requires $L_x =30a_0$, $L_{\xi}=80a_0$, $\Delta X =0.01a_0$ and $\Delta \xi=0.5a_0$.  However, for the ionic mass $M$ of $\upmu$ case, convergence required  $L_X =100a_0$, $L_{\xi}=80a_0$, $\Delta X =0.03a_0$ and $\Delta \xi=0.5a_0$. Finally, for the ED and BOA optical spectra we have used $L_{\xi} =500a_0$ and $\Delta \xi=0.1a_0$.
 
For the H$_2$ type molecules, ground state convergence is achieved for $L_X=L_{\xi}=L_x=10a_0$, $\Delta X=0.07a_0$, $\Delta \xi=0.2a_0$ and $\Delta x=0.5a_0$. To obtain the PES we have used $L_\xi=L_{x}=40a_0$ and $\Delta \xi =\Delta x =0.2a_0$. The convergence of the QMI optical spectra requires $L_X=10a_0$, $L_x=80a_0$, $L_{\xi}=35a_0$, $\Delta X=0.07a_0$, $\Delta x = 0.5a_0$ and $\Delta \xi = 0.6a_0$. Finally, for the ED and BOA optical spectra we have used $L_{\xi}=L_x=200a_0$, $\Delta \xi = \Delta x = 0.5a_0$.

Within the BOA and ED, the $X$ coordinate does not need to be discretized quantum mechanically. It is either fixed as a parameter in BOA or it changes according to the dynamic equations in ED. As a consequence, the two and three variable bare Coulomb QMI problems confined to 1D trajectories for the H$_2^+$ and H$_2$ molecules respectively, become one and two variable BOA and ED problems. These are easier to compute numerically, thus providing a more attractive alternative.

\section{Results and Discussion}\label{results:sec}

\subsection{H$_2^+$ and H$_2$ results}\label{H2results:sec}

\begin{figure}%[!ht]
\begin{center}
\includegraphics[scale=0.3]{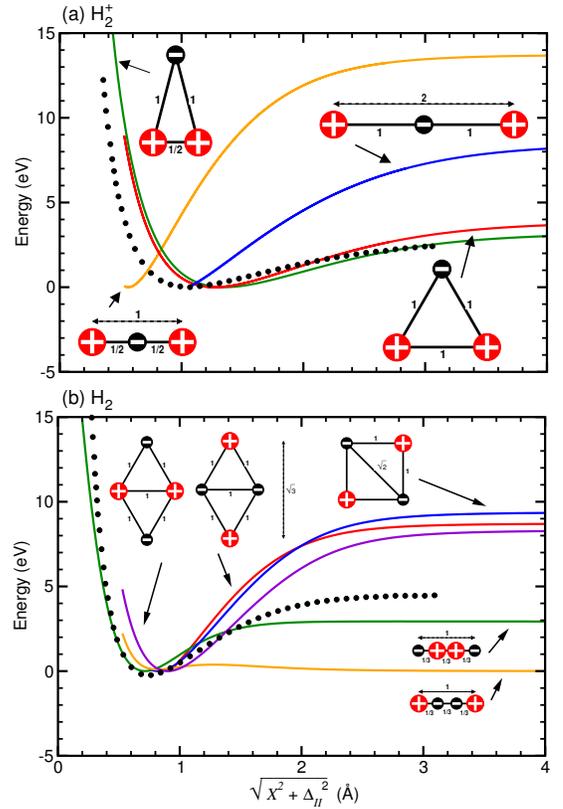}
\end{center}
\caption{
(Color online) BOA ground state PESs relative to $E_{0}(X_{\textit{eq}})$ in eV,
versus $\sqrt{X^2 + \Delta_{\textit{II}}^2}$ in $\AA$\ for (a) H$_2^+$ and (b) H$_2$ molecules for 
the $\Delta_{\textit{II}}$, $\Delta_{\textit{ee}}$ and $\Delta_{\textit{Ie}}$ configurations 
shown as insets. The 3D ground state PESs (dotted lines) have been taken from Ref.~\onlinecite{3DH2}.}
\label{BOA_PES_CASES:fig}
\end{figure}

\begin{figure}%[ht]
\begin{center}
\includegraphics[scale=0.3]{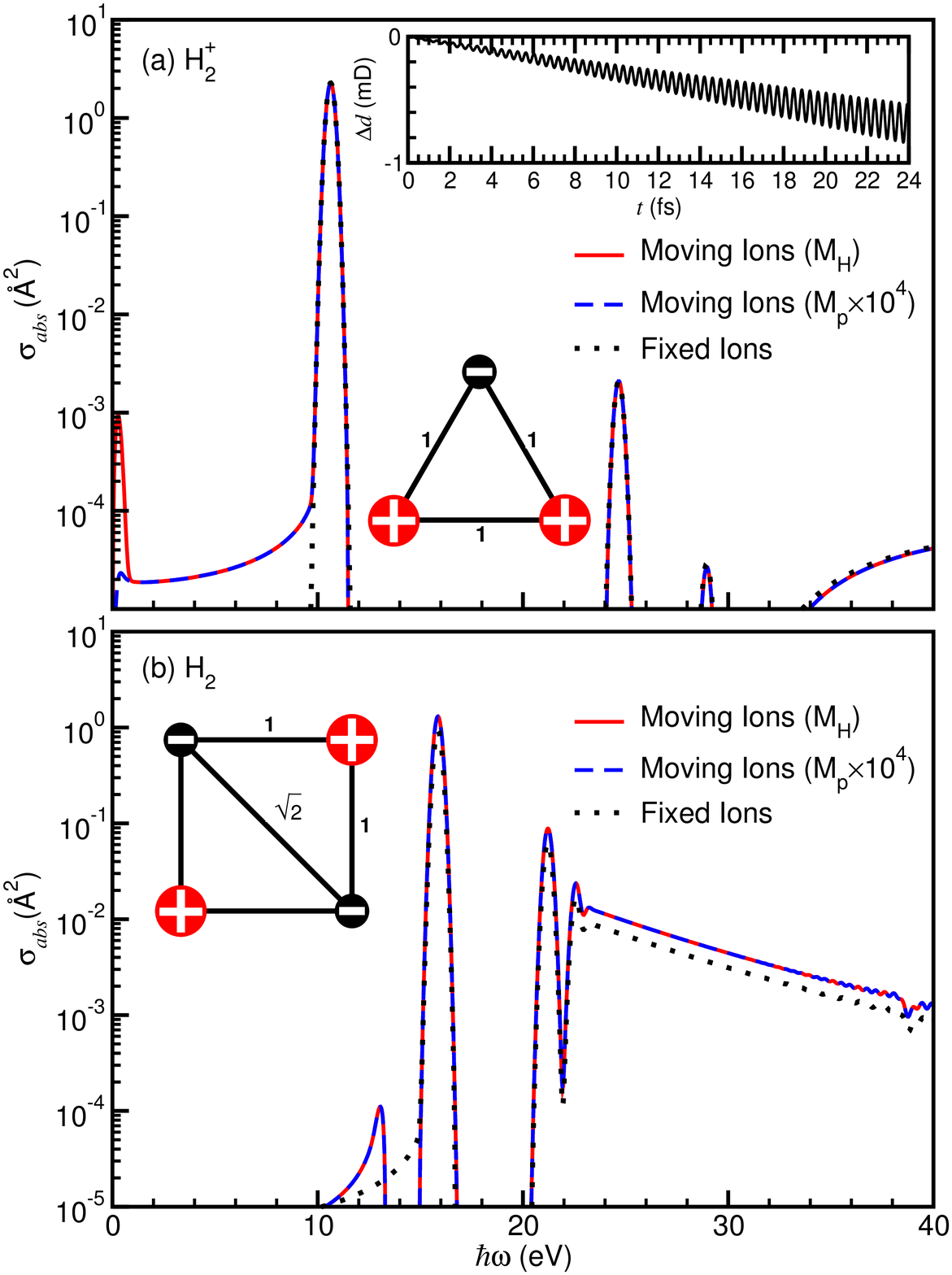}
\end{center}
\caption{
(Color online) Optical spectra for the (a) H$_2^+$ and (b) H$_2$ molecules obtained by classically 
fixing the ions to their equilibrium positions (BOA) and evolving the ions 
(ED) for masses $M_\textrm{H}$ and $M_{\textrm{p}}\times 10^4$ with minimum ionic separations (a) $\Delta_{\textit{II}}=a_0$ (b) $\Delta_{\textit{II}} = \Delta_{\textit{ee}} = \sqrt{2}a_0$ 
and electron-ion separations (a,b) $\Delta_{\textit{Ie}}=a_0$ shown as insets.  Evolution of the difference in dipole moment $\Delta d$ between ED for $M_{\textrm{H}}$ and BOA in milliDebye is shown as an inset of (a).
}
\label{ED_VS_BOA_SPECTRA:fig}
\end{figure}

In Fig.~\ref{BOA_PES_CASES:fig}, 
we show how the H$_2^+$ and H$_2$ BOA ground state PES change 
as a function of the ionic separation $\sqrt{X^2 + \Delta_{\textit{II}}^2}$ 
for each configuration shown in Fig.~\ref{EPS_VALUES:fig}.  
\begin{table}
\begin{ruledtabular}
\caption{H$_2^+$ and H$_2$ ground state PES fitted ground state energies
$E_{0}(X_{\textit{eq}})$ and positions $\sqrt{X_{\textit{eq}}^2 + \Delta_{\textit{II}}^2}$ 
obtained from a harmonic fit around $X_{\textit{eq}}$ for the configurations shown in Fig.~\ref{EPS_VALUES:fig}.}
\label{PARAMETERS}
\begin{tabular}{cccccc}
Species & $\Delta_{\textit{II}}$& $\Delta_{\textit{Ie}}$& $\Delta_{\textit{ee}} $& $E_{0}(X_{\textit{eq}})$&
$\sqrt{X_{\textit{eq}}^2 + \Delta_{\textit{II}}^2}$\\
& ($a_0$) &($a_0$) &($a_0$) & (eV) & ($\AA$)\\
\hline
\multirow{4}{*}{H$_2^+$}& 1& 0.5 & --- &
-45.757&
0.5627\\
& 0.5& 1 & --- &
-21.431&
1.3510\\
& 1& 1& --- &
-21.969&
1.2697\\
& 2 & 1 & --- &
−26.759&
1.0584\\
\hline
\multirow{4}{*}{H$_2$}
& $\frac{1}{3}$& $\frac{1}{3}$; $\frac{2}{3}$& 1 &
-60.022&
0.7146\\
& $\sqrt{3}$& 1& 1 &
-45.193&
0.9166\\
& 1& 1& $\sqrt{3}$&
-44.790&
0.9004\\
& $\sqrt{2}$& 1& $\sqrt{2}$ &
-45.856&
0.8241\\
\end{tabular}
\end{ruledtabular}
\end{table}
The PES fitted minimum energies at $X_{\textit{eq}}$, $E_{0}(X_{\textit{eq}})$ 
and positions $\sqrt{X_{\textit{eq}}^2 + \Delta_{\textit{II}}^2}$ are shown in Table \ref{PARAMETERS}
for the H$_2^+$ and H$_2$ molecules with the configurations shown in 
Fig.~\ref{EPS_VALUES:fig}.

The ground state PESs (dotted black lines in Fig.~\ref{BOA_PES_CASES:fig} and taken from Ref. \onlinecite{3DH2}) have been 
obtained by solving the stationary Schr\"{o}dinger equation in 3D using basis sets within the BOA.  
Here, the electronic and ionic positions were allowed to vary in all spatial directions.  
The overall shape of these 3D PES is reproduced qualitatively by configurations (b) and (c) for  H$_2^+$ and (g) for H$_2$ from Fig.~\ref{EPS_VALUES:fig}. 

The experimental bond lengths of H$_2^+$ and H$_2$ are $2a_0$ \cite{BOND_LENGTH_H2+}
and $\sqrt{2}a_0$ \cite{3DH2}, respectively.  The equilibrium distance is best reproduced by configuration (d) for
H$_2^+$ and (g) for H$_2$ from Fig.~\ref{EPS_VALUES:fig}. 

The Hamiltonian for configuration (g) for H$_2$ in Fig.~\ref{EPS_VALUES:fig} (b) is not invariant under electron exchange.  Yet, we still consider this configuration, as the H$_2$ configurations which are invariant under electron exchange (Fig.~\ref{EPS_VALUES:fig}(e,f,i)), yield PES that differ qualitatively from the 3D PES, as shown in Fig. \ref{BOA_PES_CASES:fig} (b).

The ions sometimes undergo a strong inter-ionic repulsion for 
larger and small $X$, depending on the initial configuration.
For the strongly repulsive configurations for small $X$, the ions are farther apart because the
repulsion between the ions is stronger than the attraction between the ions and electrons.
For the strongly attractive configurations for larger $X$,  the ions are closer together because the
repulsion between the ions is weaker than the attraction between the ions and electrons.
For the latter configurations, more energy is required to dissociate the molecule. 

For H$_2^+$, when $\Delta_{\textit{Ie}}=a_0$, the potential becomes less repulsive for small $X$ as $\Delta_{\textit{II}}$ increases.
However, when $\Delta_{\textit{Ie}}=0.5a_0$, the potential becomes strongly attractive for larger $X$.

For H$_2$, the potential becomes strongly repulsive for small $X$ for
the linear configuration (g). When $\Delta_{\textit{II}}=a_0$ and $\Delta_{\textit{Ie}} \neq a_0$, configuration (d)
in Fig.~\ref{EPS_VALUES:fig}, the ground state PES is unbound.
As the electrons are necessarily very close to each other ($\Delta_{\textit{ee}} = \frac{1}{3}a_0$) when the molecule is bound, their repulsion forces the dissociation of the H$_2$ molecule into two isolated stabler H atoms. 
For this reason we will disregard this configuration from hereon.

\begin{figure}[!t]
\begin{center}
\includegraphics[scale=0.3]{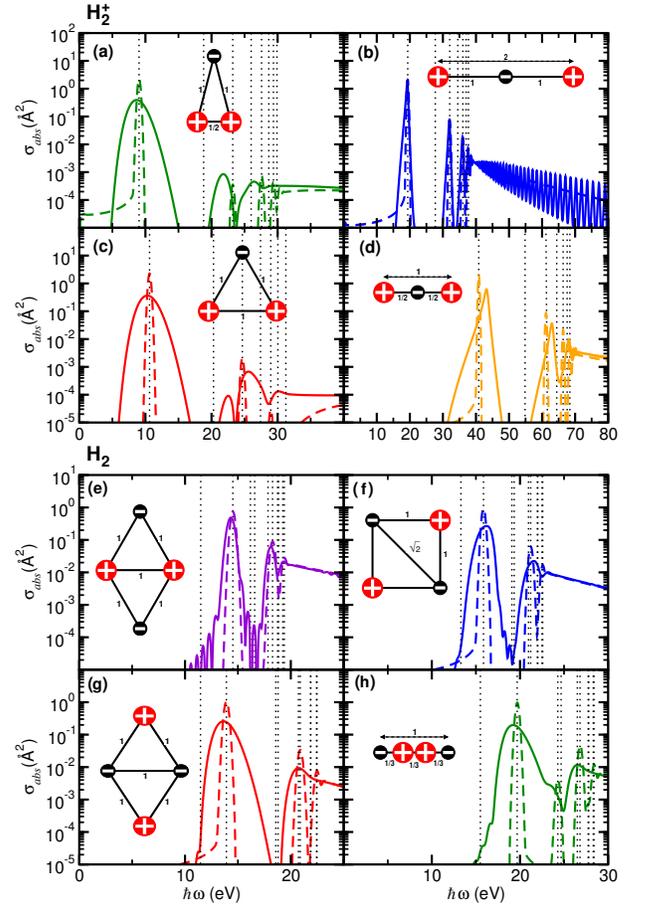}
\end{center}
\caption{
(Color online) Absorption spectra obtained from a classical BOA (dashed lines) or quantum mechanical QMI (solid lines) treatment of the ions of an (a--d) H$_2^+$  molecule with configurations (a) $\Delta_{\textit{II}} = \frac{1}{2}a_0; \Delta_{\textit{Ie}} = a_0$ (green), (b) $\Delta_{\textit{II}} = 2a_0; \Delta_{\textit{Ie}} = a_0$ (blue), (c) $\Delta_{\textit{II}} = a_0; \Delta_{\textit{Ie}} = a_0$ (red), and (d) $\Delta_{\textit{II}} = a_0; \Delta_{\textit{Ie}} = \frac{1}{2}a_0$ (orange) or an (e--h) H$_2$ molecule with configurations (e) $\Delta_{\textit{II}} = a_0; \Delta_{\textit{Ie}} = a_0; \Delta_{\textit{ee}} = \frac{1}{3}a_0$ (violet), (f) $\Delta_{\textit{II}} = \sqrt{2}a_0; \Delta_{\textit{Ie}} = a_0; \Delta_{\textit{ee}} = \sqrt{2}a_0$ (blue), (g) $\Delta_{\textit{II}} = \sqrt{3}a_0; \Delta_{\textit{Ie}} = a_0; \Delta_{\textit{ee}} = a_0$ (red), and (h) $\Delta_{\textit{II}} = \frac{1}{3}a_0; \Delta_{\textit{Ie}} = \frac{1}{3},\frac{2}{3}a_0; \Delta_{\textit{ee}} = a_0$ (green) shown as insets.  
Dotted vertical lines denote the energies $\varepsilon_i$ of the unoccupied electronic levels $\varphi_i^{X_{\textit{eq}}}(\xi)$ relative to the ground state energy $\varepsilon_0$ for each configuration at $X_{\textit{eq}}$.  
}
\label{EXACT_CHANGE_EPS:fig}
\end{figure}

In Fig.~\ref{ED_VS_BOA_SPECTRA:fig} we compare the H$_2^+$ and H$_2$ optical spectra we obtain by classically fixing and letting the ions evolve according to ED in time from $X_{\textit{eq}}$. Essentially, including the classical movement of the ions hardly changes the spectra.  However, new peaks appear before the first electronic excitation for both the H$_2^+$ and H$_2$ molecule, at 1 and 12~eV, respectively. For H$_2^+$, the new peak corresponds to the frequency of the ionic zero-point motion around $X_{\textit{eq}}$, which vanishes for large masses ($M_{\mathrm{p}}\times10^4$) because heavy ions hardly move around $X_{\textit{eq}}$. On the other hand, H$_2$'s higher energy peak does \textit{not} vanish for large $M$.  Due to its width, the ED and fixed ion spectra do not overlap completely.  We explain the origin of this peak in Sec.~\ref{data_analysis:sec}.

The inset of Fig.~\ref{ED_VS_BOA_SPECTRA:fig}(a) illustrates the time-dependent effects of zero-point motion. In the BOA the ionic center of mass $X_{\textrm{CM}_1}$ is fixed, so the electron can only oscillate about it. ED (and QMI), however, allow ionic motion, so long as the global center of mass $X_{\textrm{CM}_2}$ is conserved. The increasing difference $\Delta d$ between the ED and BOA dipole moments thus demonstrates the ions move, e.g.,  $\Delta d(24\unit{fs})\sim 0.7\unit{mD}$.

In Fig.~\ref{EXACT_CHANGE_EPS:fig}, we show how a quantum mechanical treatment of the ions (QMI) affects the optical absorption spectra for H$_2^+$ and H$_2$ molecules in the configurations of $\Delta_{\textit{II}}$, $\Delta_{\textit{ee}}$ and $\Delta_{\textit{Ie}}$ shown in Fig.~\ref{EPS_VALUES:fig}.  We see that new features emerge in the spectra when the ions are treated quantum mechanically instead of classically. The peaks are broadened, become asymmetric, and their amplitudes and energies change as a function of the initial configuration and charge of the molecule.  In particular, comparing the BOA and QMI spectra shown in Fig.~\ref{EXACT_CHANGE_EPS:fig}, we find that each peak splits into a  lower and higher energy contribution. Depending on the energy shift and amplitude of each contribution, these can appear as separate peaks or shoulders in the spectra. The shoulders are giving rise to an asymmetry that can be seen for almost every peak. These quantum features are not as strong for the neutral $\mathrm{H}_2$ homonuclear diatomic  molecule, regardless of the initial configuration. With a classical description of the ions, we do not obtain these quantum mechanical features in the optical spectra.  

Generally, we find treating the ions quantum mechanically substantially affects both the peak positions and widths in the absorption spectra for most of the configurations considered.  For inter-ionic potentials which are less repulsive (Fig.~\ref{EXACT_CHANGE_EPS:fig}(b) and (e)), the line shape of the QMI peaks is narrowed, and approaches the fixed-ion at $X_{\textit{eq}}$ limit.   For potentials which are attractive for larger $X$ (Fig.~\ref{EXACT_CHANGE_EPS:fig}(d) and (f)), all the QMI peaks are blue shifted with respect to the fixed-ion at $X_{\textit{eq}}$ spectra. In this case, the peak excitation energies are larger because more energy is required to excite these transitions.

\subsection{Mass dependency}\label{mass_dependence:sec}

To provide a quantitative analysis of the differences between a classical (BOA/ED) or quantum (QMI) treatment of the ions, we will compare the total ground state energies and the peak positions and widths in the absorption spectra as we vary the ionic mass over seven orders of magnitude.

The accuracy of the static BOA and ED calculations can be understood from a perturbation theory argument in terms of the small parameter $\kappa = (m_e/M)^{\textrm{\sfrac{1}{4}}}$, defined as the ratio between the ionic and electronic displacement \cite{BOA_VALIDITY_ONE_FOURTH}, where $X = X_{\textit{eq}} + \kappa \zeta$. We illustrate this in detail in Appendix \ref{AppendixB}.

To test the accuracy of the BOA and ED approximations, we first compare the BOA and ED ground state electron-ion eigenvalues to those obtained from QMI. We expect that the BOA and ED should be accurate around the minimum of the ground state PES, as the exact eigenvalues of the electron-ion problem can be interpreted in terms of the ionic vibrational levels for the electronic ground state PES. As discussed in Sec.~\ref{BOA_molecules:sec}, the ionic contribution comes from the ground state level of a quantum harmonic oscillator where the mass
is included via $\omega_I$.

The ground state electron-ion eigenvalue for H$_2^+$ and H$_2$ molecules whose motion is confined in one direction is given by
\begin{equation}
\varepsilon_{\textit{gs}}^{\textit{BOA}} \approx \varepsilon^{(0)} + \varepsilon^{(2)}(\kappa^2) + \bigO(\kappa^4), 
\end{equation}
where $\varepsilon^{(0)}$ and $\varepsilon^{(2)}$ correspond 
to the electronic and ionic motion eigenvalues,
and $\varepsilon^{(1)}$ and $\varepsilon^{(3)}$ are equal to zero by symmetry (see Appendix~\ref{AppendixB}).
The first order correction to the ground state energy for the full electron-ion problem using the BOA/ED is the term of fourth order in $\kappa$. 

\begin{figure}%[ht]
\begin{center}
\includegraphics[scale=0.3]{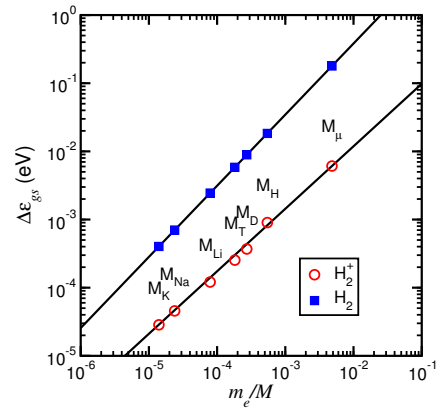}
\end{center}
\caption{
(Color online) Difference in ground state total energy between QMI and BOA/ED approaches $\Delta \varepsilon_{\textit{gs}}$ in eV versus the electron-ion mass ratio $m_e/M$ for H$_2^+$ ($\Delta_{\textit{II}}=\Delta_{\textit{Ie}}=a_0$; {\color{red}{$\medcirc$}}) and H$_2$ ($\Delta_{\textit{II}}= \Delta_{\textit{ee}} = \sqrt{2}a_0$ and $\Delta_{\textit{Ie}}=a_0$; {\color{blue}{$\blacksquare$}}).  Solid lines are a power law fit $a \left(m_e/M\right)^b$.}
\label{FIT_BOA_ERROR:fig}
\end{figure}

To check the dependence of the static ground state eigenvalue accuracy of the BOA and ED
approaches, on the electron-ion mass ratio, we use the following power law relation
\begin{equation}\label{BOA_ERROR_1D:eqn}
\varepsilon_{\textit{gs}}^{\textit{BOA/ED}} - \varepsilon_{\textit{gs}}^{\textit{QMI}} \approx a\kappa^{4b} = a {\left(\frac{m_e}{M}\right)}^b.
\end{equation}
 Note that most of the molecules used in this analysis are fictitious because we do not change the charge of the ions as explained in Sec.~\ref{model_1D:sec}, except for H, D, and T, as these have a positive electric charge of $e$. 

\begin{table}[!t]
\begin{ruledtabular}
\caption{M$_2^+$ ground state eigenvalues obtained from diagonalization of the QMI approach $\varepsilon_{\textit{gs}}^{\mathrm {QMI}}$, and ground state harmonic BOA and ED $\varepsilon_{\textit{gs}}^{ \mathrm {\textit{BOA/ED}}}$ eigenvalues obtained from a harmonic fit around the minimum of the ground state PES. We show these results for different ionic masses $M$ and $\Delta_{\textit{II}}=\Delta_{\textit{Ie}}=a_0$.}
\label{TABLE_ERRORS}
\begin{tabular}{cccc}
$M$& $\varepsilon_{\textit{gs}}^{\mathrm {QMI}}$(eV)& $\varepsilon_{\textit{gs}}^{ \mathrm {\textit{BOA/ED}}}$ (eV)& $\frac{m_e}{M}$ \\
\hline
$\upmu$&
-21.703395&
-21.697297&
0.004836\\
H&
-21.970225&
-21.969323&
0.000545\\
D&
-22.009642&
-22.009272&
0.000272\\
T&
-22.027041&
-22.026787&
0.000182\\
Li&
-22.053541&
-22.053420&
0.000079\\
Na&
-22.076626&
-22.076580&
0.000023\\
K&
-22.083189& 
-22.083160&
0.000014\\
\end{tabular}
\end{ruledtabular}
\end{table}

\begin{table}[!t]
\begin{ruledtabular}
\caption{M$_2$ ground state eigenvalues obtained by electron-ion QMI diagonalization $\varepsilon_{\textit{gs}}^{\mathrm {QMI}}$, and ground state harmonic BOA and ED $\varepsilon_{\textit{gs}}^{\mathrm {\textit{BOA/ED}}}$ vibration levels obtained from a harmonic fit around the minimum of the ground state PES. We show these results for different electron-ion mass ratios $\frac{m_e}{M}$ and $\Delta_{\textit{II}}=\sqrt{2}a_0$ and $\Delta_{\textit{Ie}}=a_0$.}
\label{TABLE_ERRORS_H2}
\begin{tabular}{cccc}
$M$&
$\varepsilon_{\textit{gs}}^{\mathrm {QMI}}$ (eV)&
$\varepsilon_{\textit{gs}}^{\mathrm {\textit{BOA/ED}}}$ (eV)&
$\frac{m_e}{M}$ \\\hline
$\upmu$&
-45.621166&
-45.440564&
0.004836\\
H&
-45.739425&
-45.722955&
0.000545\\
D&
-45.772460&
-45.763130&
0.000272\\
T&
-45.786936&
-45.779126&
0.000182\\
Li&
-45.810176&
-45.807442&
0.000079\\
Na&
-45.831319&
-45.830529&
0.000023\\
K&
-45.837552&
-45.837070&
0.000014\\
\end{tabular}
\end{ruledtabular}
\end{table}

Fitting the ground state error from Eq.~(\ref{BOA_ERROR_1D:eqn}) to the data in Tables \ref{TABLE_ERRORS} and  \ref{TABLE_ERRORS_H2}, we obtain a power law of $b\approx0.92(2)$ and $b\approx1.05(2)$ for the BOA/ED approaches, as shown in Fig.~\ref{FIT_BOA_ERROR:fig}. This means the BOA and ED energy expression gives the correct total ground state energy of the full electron-ion problem up to fourth order in $\kappa$.

\begin{figure}%[ht]
\begin{center}
\includegraphics[scale=0.3]{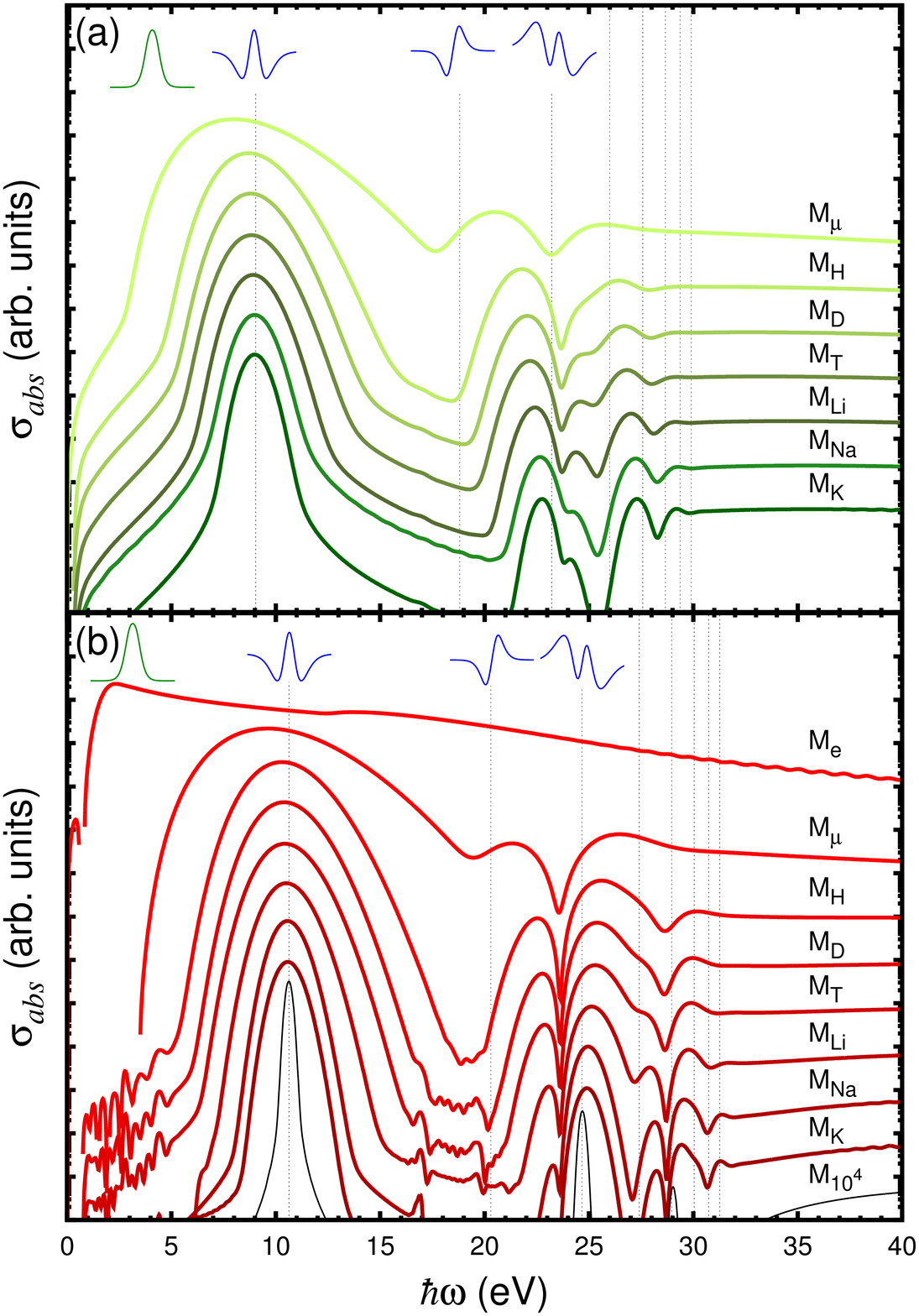}
\end{center}
\caption{
(Color online) QMI electron-ion absorption spectra for a positively charged homonuclear diatomic molecule with ionic mass $M$ of e, $\upmu$, H, D, T, Li, Na, K, or $10^4$p in the configuration (a) $\Delta_{\textit{II}}=\frac{1}{2}a_0; \Delta_{\textit{Ie}}=a_0$ or (b) $\Delta_{\textit{II}}= \Delta_{\textit{Ie}}=a_0$.  Dotted vertical lines denote the energies $\varepsilon_i$ of the unoccupied electronic levels $\varphi_i^{X_{\textit{eq}}}(\xi)$ (shown in blue as insets) relative to the energy $\varepsilon_0$ of the ground state electronic level $\varphi_0^{X_{\textit{eq}}}(\xi)$ (shown in green as insets) for each configuration.  Note that the spectra have been scaled with decreasing mass for clarity.  Portions of (b) have been adapted from Ref.~\onlinecite{Lorenzo}.
}
\label{CROSS_BOA:fig}
\end{figure}

\begin{figure}%[ht]
\begin{center}
\includegraphics[scale=0.3]{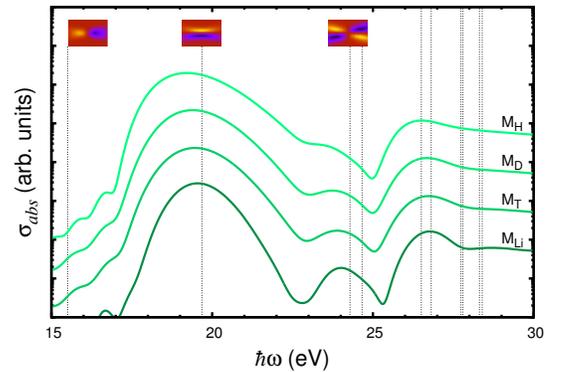}
\end{center}
\caption{
(Color online) QMI electron-ion absorption spectra for a neutral homonuclear diatomic molecule with ionic mass $M$ of H, D, T, or Li in the configuration $\Delta_{\textit{II}}=\frac{1}{3}a_0$, $\Delta_{\textit{ee}}=a_0$, and $\Delta_{\textit{Ie}}=\frac{1}{3};\frac{2}{3}a_0$.  Dotted vertical lines denote the energies $\varepsilon_i$ of the unoccupied electronic levels $\varphi_i^{X_{\textit{eq}}}(\xi,x)$ (shown as insets) relative to the ground state energy $\varepsilon_0$ for each configuration.  Note that the spectra have been scaled with decreasing mass for clarity.
}
\label{CROSS_BOA_H2_PLANAR:fig}
\end{figure}

In Figs.~\ref{CROSS_BOA:fig} and \ref{CROSS_BOA_H2_PLANAR:fig} we show how the absorption spectra depends on the mass for the H$_2^+$ and H$_2$ configurations for which the overall PES shape is closest to that from the 3D treatment in Ref.~\onlinecite{3DH2}.  Specifically, we analyze the H$_2^+$ configurations shown in Figs.~\ref{EPS_VALUES:fig}(b) and (c), and the H$_2$ configuration shown in Fig.~\ref{EPS_VALUES:fig}(g). 

In Fig.~\ref{CROSS_BOA:fig}(b) we see that in the large mass limit ($M \approx 10^4\times M_{\mathrm{p}}$), the QMI spectra exhibits even to odd transitions which are allowed by the symmetry of the electronic wavefunctions $\varphi_i(\xi)$, shown as insets. For every allowed transition in Figs.~\ref{CROSS_BOA:fig} and \ref{CROSS_BOA_H2_PLANAR:fig}, we have a red shifted and a blue shifted contribution.

The position of the first, second and fourth peaks in Fig.~\ref{CROSS_BOA:fig}(a) and (b) ($\omega_1$, $\omega_2$, and $\omega_4$) are red-shifted and the third and fifth peaks ($\omega_3$ and $\omega_5$) are blue-shifted with respect to the fixed-ion at $X_{\textit{eq}}$ spectra.  As the mass increases, all peaks tend towards the fixed-ion at $X_{\textit{eq}}$ limit. In Fig.~\ref{CROSS_BOA_H2_PLANAR:fig} all the peaks are red-shifted, although the second peak is also a classical peak as shown in Fig.~\ref{EXACT_CHANGE_EPS:fig}(h), which disappears for smaller masses.

\begin{figure}%[ht]
\begin{center}
\includegraphics[width=\columnwidth]{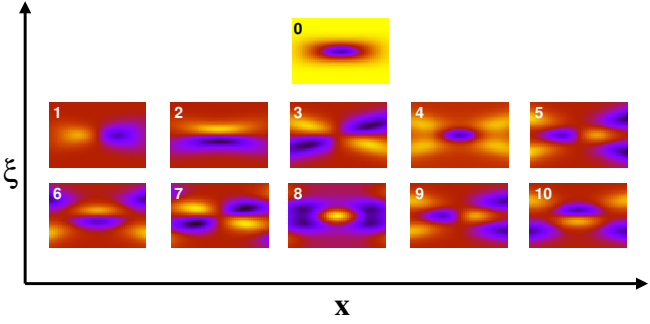}
\end{center}
\caption{
(Color online) Electronic wave functions $\varphi_i^{X_{\textit{eq}}}(\xi,x)$ for $i=0,\ldots,10$ of an H$_2$ molecule in the configuration $\Delta_{\textit{II}} = \frac{1}{3}a_0; \Delta_{\textit{ee}} = a_0$.
}
\label{WFS_H2:fig}
\end{figure}

In Fig.~\ref{WFS_H2:fig}, we show the symmetry of the occupied electronic wavefunction $\varphi_0(\xi,x)$ and the first ten unoccupied electronic wavefunctions $\varphi_i(\xi,x)$ for H$_2$.  Only transitions to unoccupied electronic wavefunctions that are even functions of $x$ and odd functions of $\xi$ should contribute to the absorption spectra by symmetry, i.e. $\langle \varphi_i(\xi,x)|\hat \xi|\varphi_0(\xi,x)\rangle > 0$.  However, Fig.~\ref{EXACT_CHANGE_EPS:fig}(h) shows there is an absorption peak in the BOA spectra for the $\varphi_0\rightarrow \varphi_3$ transition, despite $\varphi_3^{X_{\textit{eq}}}(\xi,x)$ being an odd function of $x$, as shown in Fig.~\ref{WFS_H2:fig}.  This is because the Hamiltonian for the configuration $\Delta_{\textit{II}} = \frac{1}{3}a_0$, $\Delta_{\textit{ee}} = a_0$, and $\Delta_{\textit{Ie}} = \frac{1}{3};\frac{2}{3}a_0$ is not invariant under electron exchange.

From the ED spectra in Fig.~\ref{ED_VS_BOA_SPECTRA:fig}(b), we also have an additional peak at a lower energy. When the ions are fixed, this peak is less intense than when they are allowed to evolve. In this case, the peaks' energy is given by the first excited transition ($\varphi_0\rightarrow\varphi_1$) as seen from the energy of the vertical dotted frozen ion lines in Fig.~\ref{EXACT_CHANGE_EPS:fig}(f).  However, the unoccupied wavefunction $\varphi_1(\xi,x)$ is even with respect to $\xi$ and odd with respect to $x$, as shown in Fig.~\ref{WFS_H2:fig}. This suggests such a transition should initially be parity forbidden. 

To calculate the dipole moment for H$_2$ and different $M$, we only kick our molecules along $\xi$, as shown in Sec.~\ref{spectra:sec}. When calculating the spectra in the BOA, the ionic coordinate is frozen at $X_{\textit{eq}}$ and cannot evolve in time. The electronic coordinate $x$ forms part of the integral, but can evolve in time. When we apply a kick along $\xi$, the distribution of the charge in the molecule will change with time. The electrons and ions will feel the charge distribution of the other particles. Thus, the electronic coordinate $x$ can evolve in time, although this effect is not taken into account when the dipole moment is calculated. Essentially, the ($\xi,x$) basis is rotated by the kick to a ($\xi',x'$) basis.

As $\xi$ is time dependent within ED, the $\varphi_1(\xi',x')$ rotated basis has a mixture of even and odd components in both $x$ and $\xi$, removing the parity constraint on the $\varphi_0\rightarrow\varphi_1$ transition. 

As shown in Fig.~\ref{ED_VS_BOA_SPECTRA:fig}(b), this extra parity forbidden peak is not as intense as the other peaks which are allowed by symmetry. This peak is rather weak because the electrons and ions are close to each other, but not on the same plane, for the configuration shown in Fig. \ref{EPS_VALUES:fig}(i). Additionally, as the $\varphi_0\rightarrow\varphi_1$ transition is initially forbidden by symmetry for both $x$ and $\xi$, the rotated contribution with even and odd symmetry in $x$ and $\xi$ is small.

\begin{figure}
\begin{center}
\includegraphics[width=\columnwidth]{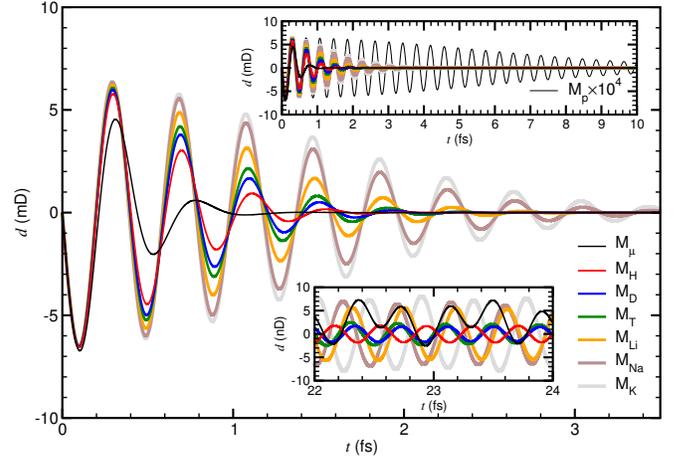}
\end{center}
\caption{(Color online) Evolution of the QMI electron-ion dipole moment $d$ in milliDebye with time $t$ after the initial ``kick'' in fs for ionic masses $M$ of $\upmu$, H, D, Ti, Li, Na, K, or (upper inset) $10^4$p of a positively charged homonuclear diatomic molecule in the configuration $\Delta_{\mathrm{II}}= \Delta_{\textit{Ie}}=a_0$. (lower inset) After 20~fs, the amplitude has decreased by a factor of one million.
}\label{Multipoles:fig}
\end{figure}

Overall, heavier ions have narrower peaks as we approach the classical limit. Figure \ref{Multipoles:fig} presents this effect in the time domain. Energy transfer between the excited electrons and the ionic system is already clearly seen after a few fs. Even in the large-mass limit ($M_p\times 10^4$) energy transfer is clearly evident. Oscillatory behavior, including beat frequencies, is still present 24~fs after the initial kick.

When the ions evolve quantum mechanically, the electrons can transfer part of their dipole moment to the ions. The amplitude of the dipole moment thus decreases at different rates depending on the ionic mass. This process will take longer as the mass of the ions increases and it becomes more difficult to displace the ions.  For very large ion masses, the interaction with the electronic motion becomes nearly elastic.  This allows the electrons to oscillate back and forth without the influence of any external ionic displacements. Since the widths of the absorption peaks are proportional to the energy transfer from the electrons to the ions, we expect the widths to scale as the electron-ion mass ratio to the one fourth, as discussed in Section~\ref{BOA:sec} and Appendix~\ref{AppendixB}. 

\subsection{Spectral lineshape}\label{data_analysis:sec}

To quantify the width and energy of the peaks in the spectra, we have employed both Gaussian 
\begin{equation}\label{gaussian:eqn}
\sum_{i=1}^{3} I_i e^{-\frac{(\omega-\omega_i)^2}{2\sigma_i^2}}
\end{equation}
and Lorentzian 
\begin{equation}\label{lorentzian:eqn}
\sum_{i=1}^{3} I_i \frac{(\Gamma_{i}/2)^2}{{(\omega-\omega_i)}^2 + (\Gamma_{i}/2)^2}.                 
\end{equation}
functions. Here $I_{i}$ is the intensity, $\omega_{i}$  the position, $\sigma_{i}$ the standard deviation, and $\Gamma_{i}$ the full width at half maximum of the first three peaks of the QMI spectra.

\begin{figure}
\begin{center}
\includegraphics[scale=0.3]{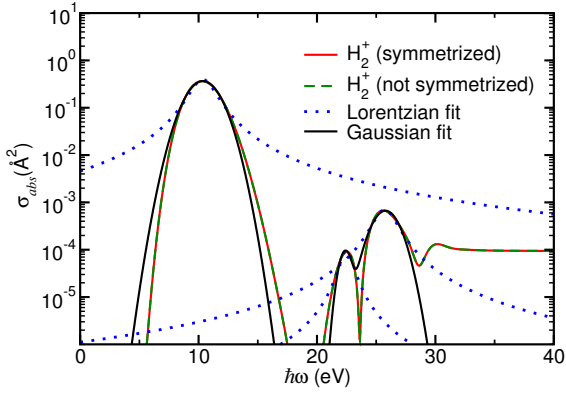}
\end{center}
\caption{
(Color online) QMI electron-ion absorption spectra for an H$_2^+$ molecule in the configuration $\Delta_{\textit{II}}=\Delta_{\textit{Ie}}=a_0$ obtained with (red solid lines) and without (green dashed lines) imposing symmetry in $X$ on the ionic wave functions. Lorentzian (blue dotted lines) and Gaussian (black solid line) fits to the first three peaks of the spectra are also provided.
}
\label{H2+_FITS:fig}
\end{figure}

From Fig.~\ref{H2+_FITS:fig}, in which we show the QMI spectra for H$_2^+$, we clearly see that the tails of the peaks of the QMI spectra are Gaussian. Moreover, the three peaks can only be fitted simultaneously with Gaussian functions, as the Lorentzian fit to the first peak decays so slowly that the second and third peaks are completely obscured. Furthermore, the ionic wave packet on the ground state PES is a solution of a harmonic eigenvalue problem and thus should have a Gaussian line shape.   This means the spectral line shape arises from the shape of the PES, rather than the coupling between ionic vibrations of the molecule. 

Note that the width of the fixed-ion at $X_{\textit{eq}}$ spectra in Fig.~\ref{ED_VS_BOA_SPECTRA:fig} is due to the artificial damping introduced in the spectra. The electronic transitions should be delta-like functions, but are convoluted with a Gaussian function to plot the spectra (see Eqs.~(\ref{FT:eqn}) and (\ref{THIRD_ORDER_POL:eqn})). However, the widths in the QMI spectra are physical, and the Gaussian line shape is due to the electron-electron coupling via the ionic displacements.

To ensure that the optical spectra we obtain is only affected by the external perturbation $K$, we have used a symmeterized initial wavefunction 
\begin{equation}
\psi_{\textit{symm}}(X,\xi) = \frac{\psi(X,\xi) + \psi(-X, \xi)}{\sqrt{2}}.
\end{equation}
Thus, we always excite from a ground state which is symmetric in the ionic coordinate. In Fig.~\ref{H2+_FITS:fig}, we show that symmetrizing the wavefunction does not change the calculated optical absorption spectrum. This means that we already obtain a nearly symmetric ground state starting configuration from the stationary Schr\"{o}dinger equation (\ref{MB_indep:eqn}). However, the data shown in Fig.~\ref{CROSS_BOA:fig}(b) has been symmetrized for every $M$.

\subsection{Model}\label{MODEL:sec}

To explain why a quantum treatment of the ions has such a strong effect on the absorption spectra for H$_2^+$,  we propose a simple two level model.  Using this model, we will show how the observed QMI spectral peaks and widths can be extracted from the electronic BOA eigenenergies $\varepsilon_i$ at equilibrium $X_{\textit{eq}}$ of the ground state through the electron-ion mass ratio $m_e/M$.  

When an external kick is applied, a charge separation is induced in the molecule which will oscillate back and forth with time.  As discussed in Sec.~\ref{spectra:sec}, the applied kick is simply a transformation of the ground state wavefunctions, through the application of a phase factor $e^{-i K\xi}$, to eigenstates of the system with momentum $K$ along the direction of motion.  This leads to a 
transition dipole moment between an initial and a final electronic state.

The electronic wavefunctions with even indices $\varphi_{2i}$ are even functions of $\xi$, while the electronic wavefunctions with odd indices $\varphi_{2i+1}$ are odd functions of $\xi$.  This parity of the electronic wavefunctions means that the transition dipole moment is zero for transitions from the ground state to even unoccupied states, i.e., $\langle \varphi_{2i+2}|\hat \xi | \varphi_0\rangle = 0$.  Essentially, %even-even 
optical transitions $\varphi_0\rightarrow\varphi_{2i+2}$ are forbidden so long as $\varphi_{2i+2}$ is an even function of $\xi$.  %as the parity of the electronic wavefunctions is conserved.  

This is the case when the ions are treated classically.  In fact, Figs.~\ref{EXACT_CHANGE_EPS:fig}(a--d) clearly show that the peaks in the absorption spectra obtained from a classical BOA treatment are always aligned with the energies of odd-parity unoccupied electronic levels $\varepsilon_{2i+1}$, i.e., the Franck-Condon transitions $\varphi_0\rightarrow\varphi_{2i+1}$.  

When the ions are treated quantum mechanically, every allowed transition is split into red and blue shifted contributions, with the shifts increasing as the mass decreases.  The level splitting we observe in Fig.~\ref{CROSS_BOA:fig} is reminiscent of level hybridization.

This motivates us to employ a simple two-level model \cite{Hammer1995211, MowbrayJPCL2013} to describe the energies and widths of the QMI peaks.

To do so, for each odd-parity unoccupied electronic level $\varphi_{2i+1}$ at $\varepsilon_{2i+1}$, we artificially introduce a level at $\widetilde{\varepsilon}_{2i+1}$ to which it couples.

As mentioned in Section~\ref{BOA:sec} and Appendix~\ref{AppendixB}, the ratio between the vibrational and electronic energies, $E_{\textit{vib}}/E_{\textit{elec}}$, scales as the square of the ratio between the ionic and electronic displacement $(\delta/a_0)^2$ \cite{BOA_VALIDITY_ONE_FOURTH}. This means the ionic displacement scales as the electron-ion mass ratio to the one fourth $\delta \approx (m_e/M)^{\textrm{\sfrac{1}{4}}}$. Since the probability of coupling is directly related to the quantum ionic displacement, we expect the coupling between the energy levels to scale as $\delta \approx (m_e/M)^{\textrm{\sfrac{1}{4}}}$. Further, the width of the peaks in the absorption spectra should also be related to the ionic displacement. We thus assume the coupling between the energy levels is proportional to the ionic displacement $\delta \sim M^{-\sfrac{1}{4}}$, i.e, $\alpha M^{-\sfrac{1}{4}}$ where $\alpha$ is the constant of proportionality. 

The resulting two-level Hamiltonian $\left[\begin{array}{cc}
\varepsilon_{2i+1} & \frac{\alpha}{M^{\textrm{\sfrac{1}{4}}}}\\
\frac{\alpha}{M^{\textrm{\sfrac{1}{4}}}} & \widetilde{\varepsilon}_{2i+1}
\end{array}\right]$ has a solution for the $\varepsilon_{2i+1}$ allowed peak's energy of
\begin{equation}
\omega_{2i, 2i+1} \approx \frac{\varepsilon_{2i+1} + \widetilde{\varepsilon}_{2i+1}}{2} - \sqrt{\left(\frac{\varepsilon_{2i+1} - \widetilde{\varepsilon}_{2i+1}}{2}\right)^2 + \left(\frac{\alpha}{M^{\textrm{\sfrac{1}{4}}}}\right)^2}\label{omega1:eqn}.
\end{equation}

This two-level model yields two peaks that are lower and higher in energy, through level repulsion. As the mass decreases, ionic displacements become larger.  This leads to a greater coupling. The coupling between the energy levels will be larger when the mass decreases and the energy separation between the two coupling energy levels will also increase.

\begin{figure*}[!t]
\begin{center}
\includegraphics[scale=0.3]{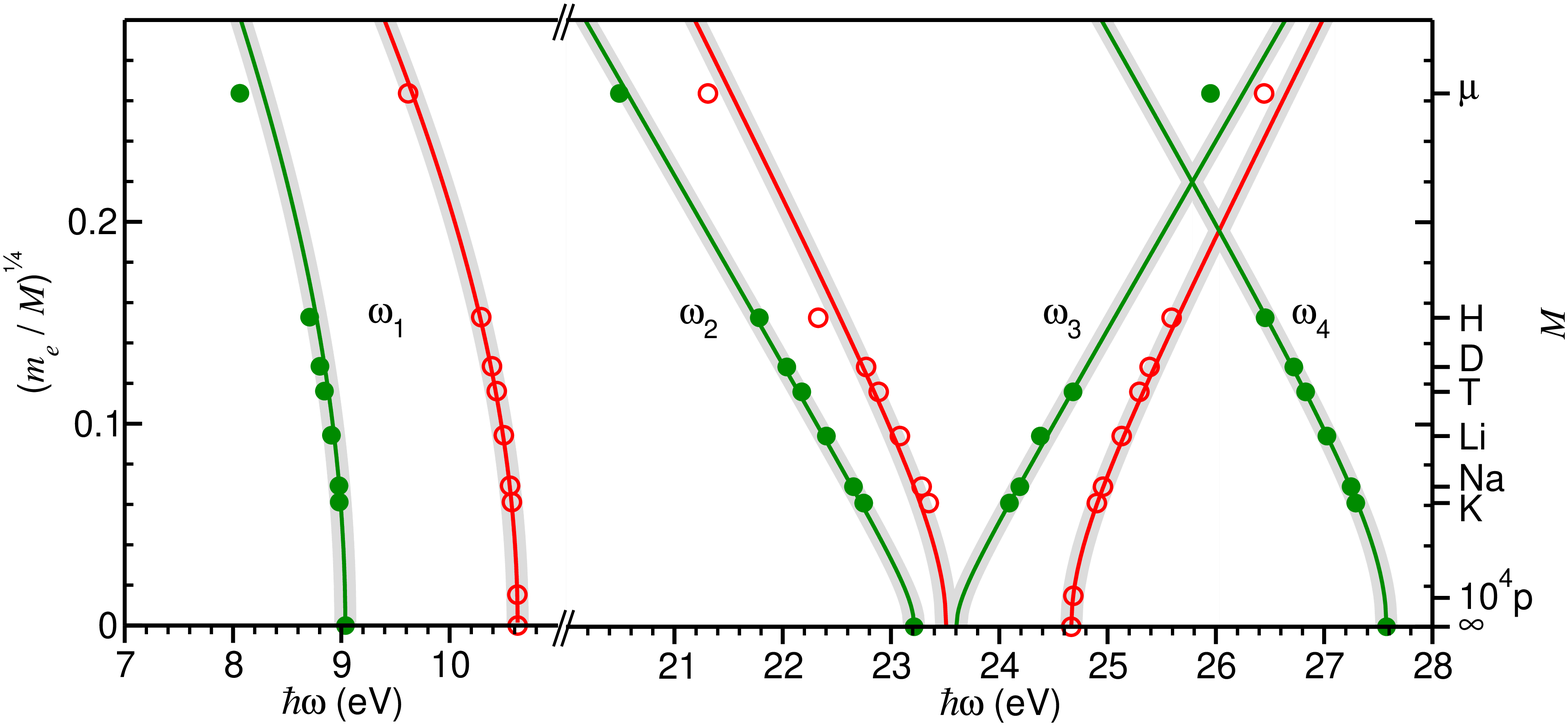}
\end{center}
\caption{
(Color online) Two-level model fits to the first four peaks in the QMI electron-ion absorption spectra $\omega_{i}$ for a positively charged diatomic molecule with ionic mass $M$ in the configuration $\Delta_{\textit{Ie}} = a_0$ and $\Delta_{\textit{II}} = \frac{1}{2}a_0$ ({\color{Green}{$\medbullet$}}) or $\Delta_{\textit{II}} = a_0$ ({\color{red}{$\medcirc$}}).  Level coupling has the form $\alpha (m_e/M)^{\textrm{\sfrac{1}{4}}}$, and the decoupled levels are obtained from the ground state electronic eigenenergies $\varepsilon_i$.  Gray regions denote a $\pm 0.1$~eV estimated accuracy.
}
\label{H2+_ALL_PEAKS:fig}
\end{figure*}

 \begin{figure}[!t]
\begin{center}
\includegraphics[scale=0.3]{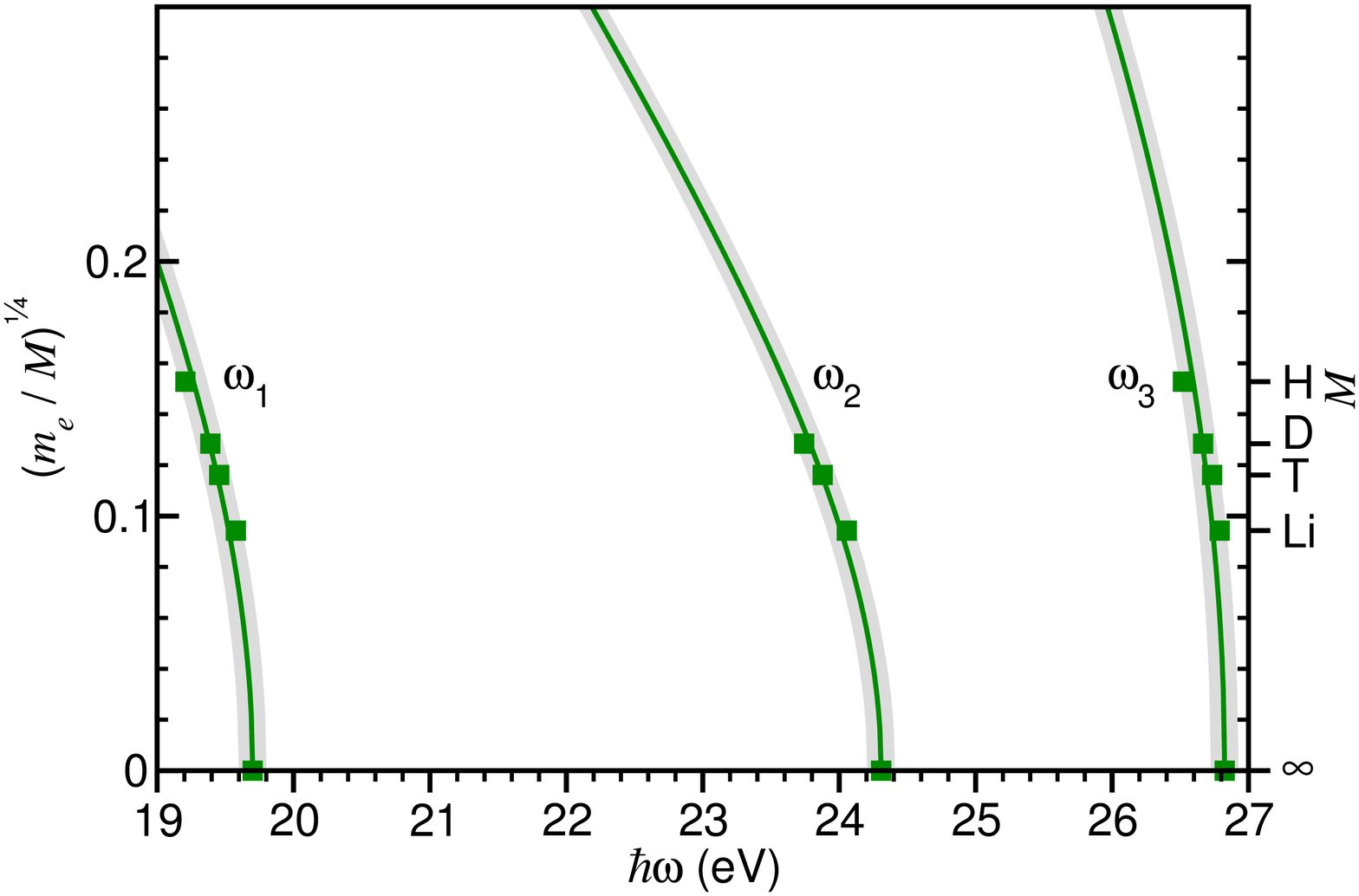}
\end{center}
\caption{
(Color online) Two-level model fits to the first three peaks in the QMI electron-ion absorption spectra $\omega_{i}$ for a neutral diatomic molecule with ionic mass $M$ in the configuration $\Delta_{\textit{Ie}} = \frac{1}{3};\frac{2}{3}a_0$, $\Delta_{\textit{II}} = \frac{1}{3}a_0$, and $\Delta_{\textit{ee}} = a_0$ ({\color{Green}{$\blacksquare$}}).  Level coupling has the form $\alpha (m_e/M)^{\textrm{\sfrac{1}{4}}}$, and the decoupled levels are obtained from the ground state electronic eigenenergies $\varepsilon_i$.  Gray regions denote a $\pm 0.1$~eV estimated accuracy.
}
\label{PLANAR_H2_ALL_PEAKS:fig}
\end{figure}

\begin{figure}%[ht]
\begin{center}
\includegraphics[scale=0.3]{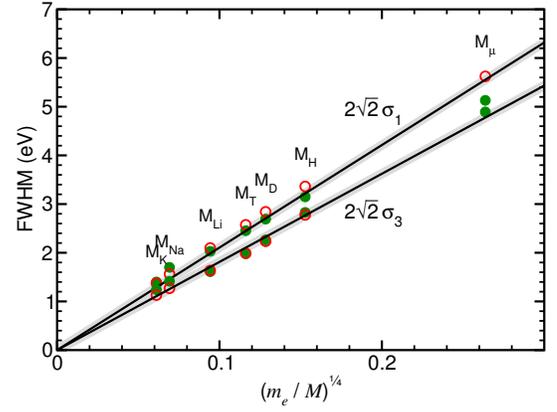}
\end{center}
\caption{
(Color online) Full width at half maximum (FWHM) of the Gaussian fits ($\mathrm{FWHM}=2\sqrt{2}\sigma_i$) to the first and third peaks of the absorption spectra for a positively charged homonuclear diatomic molecule with ionic mass $M$ of $\upmu$, H, D, T, Li, Na, or K versus the fourth root of the electron-ion mass ratio $(m_e/M)^{\mathrm{\textrm{\sfrac{1}{4}}}}$ in the configuration $\Delta_{\textit{Ie}} = a_0$ and $\Delta_{\textit{II}} = \frac{1}{2}a_0$ ({\color{Green}{$\medbullet$}}) or $\Delta_{\textit{II}} = a_0$ ({\color{red}{$\medcirc$}}).  Black lines are linear fits to each peak for both configurations. Gray regions denote a $\pm 0.1$~eV estimated accuracy.
}
\label{WIDTH_H2+:fig}
\end{figure}

In Figs.~\ref{H2+_ALL_PEAKS:fig} and \ref{PLANAR_H2_ALL_PEAKS:fig} we use the two-level model to fit the calculated QMI peaks for various ionic masses $M$.  Specifically, we employ Eqs.~(\ref{omega1:eqn}) to fit $\omega_1$, $\omega_2,$ and $\omega_3$ for H$_2^+$ and H$_2$.  In general, the calculated peak positions are within $0.1$~eV of the two-level model fit, which is also the expected accuracy of such calculations.  In each case, we find the coupling between the transitions has a constant of proportionality of $\alpha \approx 11$~eV.  Furthermore, the artificial level $\widetilde{\varepsilon}_{3} \approx 23.5$~eV for both configurations of H$_2^+$. 

For H$_2^+$, we find the first peak depends on $M^{-\textrm{\sfrac{1}{2}}}$, since \(\alpha^2/M^{\textrm{\sfrac{1}{2}}} \ll \widetilde{\varepsilon}_{1} - \varepsilon_1\approx\) 9.7~eV. In other words,
\begin{equation}
\omega_1 \approx \varepsilon_1 - \frac{\alpha^2/M^{\textrm{\sfrac{1}{2}}}}{\widetilde{\varepsilon}_{1} - \varepsilon_1}\label{approxomega1:eqn},
\end{equation}
where $\widetilde{\varepsilon}_1 \approx 20$~eV.

For the second and third peaks, so long as $\varepsilon_3-\widetilde{\varepsilon}_3 \ll 2\alpha/M^{\sfrac{1}{4}}$, we may further approximate the peaks by  
\begin{equation}
\begin{aligned}
\omega_{2,3}&\approx \frac{\varepsilon_3+\widetilde{\varepsilon}_{3}}{2} \pm \left[\frac{\alpha}{M^{\textrm{\sfrac{1}{4}}}} + \frac{(\varepsilon_3 - \widetilde{\varepsilon}_3)^2M^{\sfrac{1}{4}}}{8\alpha}\right],
\\&\approx \frac{\varepsilon_3+\widetilde{\varepsilon}_{3}}{2} \pm \frac{\alpha}{M^{\textrm{\sfrac{1}{4}}}}.
\end{aligned}
\end{equation}
As we see in Fig.~\ref{H2+_ALL_PEAKS:fig}, this is indeed the case for the second and third peaks in the absorption spectra, \(\omega_2\) and \(\omega_3\), of H$_2^+$ in the configuration $\Delta_{\textit{Ie}} = a_0$ and $\Delta_{\textit{II}} = \frac{1}{2}a_0$, as $\varepsilon_3 - \widetilde{\varepsilon}_{3} \approx 0.4$~eV.

Essentially, all the peak positions in the QMI spectra are fit using only two parameters, the artificial level's energy $\widetilde{\varepsilon}_{2i+1}$, and the coupling constant $\alpha$.  However,  we are not always able to decouple the two peaks' red and blue shifted contributions because of their overlap due to their finite width. This is particularly true for H$_2$.  As a result, we have fewer data points for the H$_2$ peaks, as shown in Fig.~\ref{PLANAR_H2_ALL_PEAKS:fig}, reducing the reliability of the fit to Eq. \ref{omega1:eqn}.

In Fig.~\ref{WIDTH_H2+:fig}, we also show that the width of the first and third peaks for the configurations
shown in Fig.~\ref{EPS_VALUES:fig} (b) and (c) scale as the electron-ion mass ratio to the one fourth, 
i.e. $\textrm{FWHM} \approx (m_e/M)^{\textrm{\sfrac{1}{4}}}$, as expected from our model.  
The FWHM for the first peak has a larger constant of proportionality than the third peak, 
but the widths for both configurations may be fit simultaneously.  
Altogether, this demonstrates the predictive power of the simple two-level model
for describing the QMI spectra as a function of the ionic mass.

\section{Conclusions}\label{conclusions:sec}

We have shown that additional features may appear in the linear response spectra of charged H$_2^+$ and neutral H$_2$ homonuclear diatomic molecules when the ionic motion is described quantum mechanically. Such features are strongly dependent on the molecules' configuration, i.e., the shape of the PES. The widely used classical ionic motion BOA and ED approaches fail to describe such features.  We also demonstrate that these features may be understood using a predictive two-level model.  These results demonstrate how for light atoms, the quantum nature of the ions may play an important role when describing absorption processes.

\section{Acknowledgements}\label{acknowledgements:sec}

The authors thank Angel Rubio, Stefan Kurth and Lorenzo Stella for useful discussions. We acknowledge financial support from the European Research Council Advanced Grant DYNamo (ERC-2010-AdG-267374), Spanish Grants (FIS2013-46159-C3-1-P  and PIB2010US-00652), and Grupo Consolidado UPV/EHU del Gobierno Vasco (IT578-13). A. C.-U. acknowledges financial support from the Departamento de Educaci\'on, Universidades e Investigaci\'on del Gobierno Vasco (Ref. BFI-2011-26) and DIPC.

\appendix

\section{Center of mass transformation}
\label{AppendixA}

Here we present in detail the coordinate transformations applied in our description of a homonuclear diatomic molecule whose electronic and ionic motion has been confined to one direction.  First, we perform a center of mass transformation of the two ionic coordinates $X_1$ and $X_2$
\begin{equation}\label{RA:eqn}
\begin{aligned}
 X_{\mathrm{CM}_1} &=  \frac{X_1+X_2}{2} ; 
 &V_{\mathrm{CM}_1} =  \frac{V_1+V_2}{2},\\
 X &= X_2 - X_1 ; &V_X = V_2 - V_1,\\
\end{aligned}
\end{equation}
where $X_{\mathrm{CM}_1}$ is the center of mass coordinate of the ions and $X$ is the distance
between the ions. Here the velocities are the time derivatives of the positions.

\subsection{Positively charged homonuclear diatomic molecule}
For a positively charged homonuclear diatomic molecule with one electron, the electronic coordinate and velocity are simply
\begin{equation}
\begin{aligned}\label{x_def:eqn}
x &= x_1 ;& v &= v_1.
\end{aligned}
\end{equation}
Next, we perform a global center of mass transformation of the center of ionic mass and electronic coordinates $X_{\mathrm{CM}_1}$ and $x$, keeping the ionic separation $X$ fixed
\begin{equation}\label{MU:eqn}
\begin{aligned}
X_{\mathrm{\mathrm{CM}_2}} &= \frac{2M X_{\mathrm{CM}_1}+ x}{2M + 1} ;&
V_{\mathrm{\mathrm{CM}_2}} &= \frac{2M V_{\mathrm{CM}_1} + v}{2M + 1}, \\
\xi &= x - X_{\mathrm{CM}_1} ;& V_{\xi} &= v - V_{\mathrm{CM}_1},
\end{aligned}
\end{equation}
where $X_{\mathrm{CM}_2}$ is the global center of mass coordinate and $\xi$ is the distance between the electron $x$ and the ionic center of mass $X_{\mathrm{CM}_1}$. Here the velocities are the time derivatives of the positions.

By substituting Eqs.~(\ref{RA:eqn}), (\ref{x_def:eqn}), and (\ref{MU:eqn}) into Eq.~(\ref{E_ini:eqn}) we obtain for the classical energy of a positively charged homonuclear diatomic molecule
\begin{equation}\label{E_nearly_total:eqn}
\begin{split}
E &= \frac{1}{2}(2M+1) V_{\mathrm{CM}_2}^2 
  + \frac{1}{2} \mu_p V_X^{2}  +  
  \frac{1}{2} \mu_e V_{\xi}^2 \\  
  &- \frac{1}{\sqrt{(\frac{X}{2} +\xi)^2 + \Delta_{\textit{Ie}}^2}} 
  -\frac{1}{\sqrt{(\frac{X}{2} -\xi)^2 + \Delta_{\textit{Ie}}^2}} + 
  \frac{1}{\sqrt{X^2 + \Delta_{\textit{II}}^2}},
\end{split}
\end{equation}
where $\mu_e$ is the reduced mass of the two ions plus electron system and $\mu_p$ is the reduced mass of the two ions
\begin{equation}\label{reduced_mass_electron:eqn}
\begin{aligned}
\mu_e &= \frac{2M}{2M+1};& \mu_p &= \frac{M}{2}.
\end{aligned}
\end{equation}

If we rewrite Eq.~(\ref{E_nearly_total:eqn}) in terms of the momenta given by
\begin{equation}\label{Momentum_quantum:eqn}
\begin{aligned}
\hat P_{X_{\mathrm{CM}_2}} &= (2M+1)V_{\mathrm{CM}_2} = -i \frac {\partial}{\partial X_{\mathrm{CM}_2}},\\
\hat P_X &= \mu_p V_X = -i \frac {\partial}{\partial X},\\
\hat P_\xi &= \mu_e V_{\xi} = -i \frac {\partial}{\partial \xi},
\end{aligned}
\end{equation}
we obtain the two-body Hamiltonian in Eq.~(\ref{quantum_H:eqn}).

\subsection{Neutral homonuclear diatomic molecule}
For a neutral homonuclear diatomic molecule with two electrons,  we also perform a center of mass transformation of the two electronic coordinates $x_1$ and $x_2$
\begin{equation}\label{JAJA:eqn}
\begin{aligned}
 x_{\mathrm{CM}_1} &= \frac{x_1+x_2}{2} ;& 
 v_{\mathrm{CM}_1} &= \frac{v_1+v_2}{2},\\
 x &= x_2 - x_1 ;& v &= v_2 - v_1,
\end{aligned}
\end{equation}
where $x_{\mathrm{CM}_1}$ is the center of mass coordinate of the electrons and $x$ is the distance between the electrons. Here the velocities are the time derivatives of the positions.

We now perform a global center of mass transformation of the two ionic and electronic center of mass coordinates coordinates $X_{\mathrm{CM}_1}$ and $x_{\mathrm{CM}_1}$, keeping the ionic and electronic separations $X$ and $x$ fixed
\begin{equation}\label{JUJU:eqn}
\begin{aligned}
X_{\mathrm{CM}_2} &= \frac{2M X_{\mathrm{CM}_1} + 2 x_{\mathrm{CM}_1}}{2M + 2} ;&
V_{\mathrm{CM}_2} &= \frac{2M V_{\mathrm{CM}_1} + 2 v_{\mathrm{CM}_1}}{2M + 2},\\
\xi &= x_{\mathrm{CM}_1} - X_{\mathrm{CM}_1} ;& 
V_{\xi} &= v_{\mathrm{CM}_1} - V_{\mathrm{CM}_1},
\end{aligned}
\end{equation}
where $X_{\mathrm{CM}_2}$ is the global center of mass coordinate and $\xi$ is the distance between $X_{\mathrm{CM}_1}$ and $x_{\mathrm{CM}_1}$. Here the velocities are the time derivatives of the positions.

By substituting Eqs.~(\ref{RA:eqn}), (\ref{JAJA:eqn}), and (\ref{JUJU:eqn}) into Eq.~(\ref{E_ini_hydrogen:eqn}) we obtain for the classical energy of a neutral homonuclear diatomic molecule 
\begin{equation}\label{E_total_hydrogen:eqn}
\begin{split}
E =& \frac{1}{2}(2M+2) V_{\mathrm{CM}_2}^2 
  + \frac{1}{2} \mu_p V_X^{2}  +  
  \frac{1}{2} \mu_{ep} V_{\xi}^2 
  + \frac{1}{2} \widetilde{\mu}_e v^{2} \\  
  &-\frac{1}{\sqrt{(\frac{X}{2} - \frac{x}{2} + \xi)^2 + \Delta_{\textit{Ie}}^2}}
  - \frac{1}{\sqrt{(\frac{X}{2} - \frac{x}{2} - \xi)^2 + \Delta_{\textit{Ie}}^2}} \\
  &- \frac{1}{\sqrt{(\frac{X}{2} + \frac{x}{2} + \xi)^2 + \Delta_{\textit{Ie}}^2}}
  - \frac{1}{\sqrt{(\frac{X}{2} + \frac{x}{2} - \xi)^2 + \Delta_{\textit{Ie}}^2}} \\
  &+ \frac{1}{\sqrt{X^2 + \Delta_{\textit{II}}^2}} 
  + \frac{1}{\sqrt{x^2 + \Delta_{\textit{ee}}^2}}\;,
\end{split}
\end{equation}
where $\mu_p$ is the reduced mass of the two ions, $\mu_{ep}$ is the reduced mass of the two ions plus two electron system, and $\widetilde{\mu}_e$ is the reduced mass of the two electrons
\begin{equation}\label{reduced_mass_electron_proton_hydrogen:eqn}
\begin{aligned}
\mu_p &= \frac{M}{2};& \mu_{ep} &= \frac{2M}{1+M};& \widetilde{\mu}_e &= \frac{1}{2}.
\end{aligned}
\end{equation}

If we rewrite Eq.~(\ref{E_total_hydrogen:eqn}) in terms of the momenta given by
\begin{equation}\label{Momentum_quantum_hydrogen:eqn}
\begin{aligned}
\hat P_{X_{\mathrm{CM}_2}} &= (2M+2)V_{\mathrm{CM}_2} = -i \frac {\partial}{\partial X_{\mathrm{CM}_2}},\\
\hat P_X &= \mu_p V_X = -i \frac {\partial}{\partial X},\\
\hat P_\xi &= \mu_{ep} V_{\xi} = -i \frac {\partial}{\partial \xi},\\
\hat P_x &= \widetilde{\mu_{e}} v = -i \frac {\partial}{\partial x},
\end{aligned}
\end{equation}
we obtain the three-body Hamiltonian ($X$,$x$,$\xi$) in Eq.~(\ref{H_int:eqn}).

\section{Accuracy of the BOA}
\label{AppendixB}

Here we provide a detailed analysis of the accuracy of the Born-Oppenheimer Approximation (BOA) \cite{BOA_ORIGINAL} for the case of a homonuclear diatomic molecule whose electronic and ionic motion is confined to one direction.  In general, the ratio of vibrational to electronic energies, $E_{\textit{vib}}$ to $E_{\textit{elec}}$ depends on the electron-ion mass ratio $m_e/M$ as \cite{BOOK_RATIO}
\begin{equation}
\frac{E_{\textit{vib}}}{E_{\textit{elec}}} \approx \sqrt{\frac{m_e}{M}} \approx \frac{\delta^2}{a_0^2},
\end{equation}
where $\delta$ is the length scale of vibrational motion, and $a_0$ is the length scale of electronic motion, i.e., the Bohr radius.  This means the ratio of ionic to electronic motion is of the order $\delta/a_0 \approx (m_e/M)^{\sfrac{1}{4}}$.  With this in mind, we may expand the Hamiltonian in Eq.~(\ref{H_int:eqn}) as a function of the small parameter $\kappa \equiv (m_e/M)^{\sfrac{1}{4}}$ \cite{BOA_VALIDITY_ONE_FOURTH} to third order as follows:
\begin{equation}\label{H_i_H2P_BOA_expanded:eqn}
\begin{split}
\hat H(X_{\textit{eq}} + \kappa \zeta,x,\xi) \approx& 
-  \frac{1}{2\widetilde{\mu}_e} \frac{\partial^2}{\partial{x}^2} 
  - \frac{1}{2\mu_{ep}} \frac{\partial^2}{\partial{\xi}^2} +
  V(X_{\textit{eq}},x,\xi)\\
& + \kappa \left.\frac{\partial}{\partial X} V(X,x,\xi) \right|_{X=X_{\textit{eq}}} \zeta \\
 &  -  \kappa^2 \frac{\partial^2}{\partial{\zeta}^2} 
+ \frac{1}{2!}\kappa^2 \left.\frac{\partial^2}{\partial X^2} V(X,x,\xi) \right|_{X=X_{\textit{eq}}} \zeta^2 \\
 &+ \frac{1}{3!}\kappa^3 \left.\frac{\partial^3}{\partial X^3} V(X,x,\xi) \right|_{X=X_{\textit{eq}}} \zeta^3 
 + \bigO(\kappa^4).
\end{split}
\end{equation}
Sorting the Hamiltonian in different powers of $\kappa$, i.e., 
\begin{equation}
\hat{H}(X_{\textit{eq}} + \kappa \zeta,x,\xi) \approx \hat H^{(0)} + \kappa \hat H^{(1)} + \kappa^2 \hat H^{(2)} + \kappa^3 \hat H^{(3)},
\end{equation}
 we obtain:
\begin{equation}
\begin{split}
\hat H^{(0)} &= - \frac{1}{2\widetilde{\mu}_e} \frac{\partial^2}{\partial{x}^2} - \frac{1}{2\mu_{ep}} \frac{\partial^2}{\partial{\xi}^2} + V(X_{\textit{eq}},x,\xi).\\
\hat H^{(1)} &= \left.\frac{\partial}{\partial X} V(X,x,\xi) \right|_{X=X_{\textit{eq}}} \zeta,\\
\hat H^{(2)} &= - \frac{\partial^2}{\partial{\zeta}^2} + \frac{1}{2!} \left.\frac{\partial^2}{\partial X^2} V(X,x,\xi) \right|_{X=X_{\textit{eq}}} \zeta^2,\\
\hat H^{(3)} &= \frac{1}{3!} \left.\frac{\partial^3}{\partial X^3} V(X,x,\xi) \right|_{X=X_{\textit{eq}}} \zeta^3.
\end{split}
\end{equation}

Expanding the time-independent Schr\"{o}dinger equation (\ref{MB_indep:eqn}) in powers of $\kappa$ to the third order, we obtain:
\begin{equation}\label{order_kappa:eqn}
\sum_{n=0}^{3}(\kappa^n \hat H^{(n)})[\kappa^n \psi^{(n)}]  = \sum_{n=0}^{3}(\kappa^n \varepsilon^{(n)}) [\kappa^n \psi^{(n)}].
\end{equation}

Decomposing Eq.~(\ref{order_kappa:eqn}) in terms of $\kappa$, we find
\begin{eqnarray}
\bigO(\kappa^0):&& \hat H^{(0)} | \psi^{(0)} \rangle = \varepsilon^{(0)} | \psi^{(0)} \rangle,\\
\label{1_order:eqn}
\bigO(\kappa^1):&& \hat H^{(0)} | \psi^{(1)} \rangle + \hat H^{(1)} | \psi^{(0)} \rangle = \varepsilon^{(0)} | \psi^{(1)} \rangle + \varepsilon^{(1)} | \psi^{(0)} \rangle,\\
\bigO(\kappa^2): && \hat H^{(0)} | \psi^{(2)} \rangle + \hat H^{(1)} | \psi^{(1)} \rangle + \hat H^{(2)} | \psi^{(0)} \rangle \nonumber\\
&&= \varepsilon^{(0)} | \psi^{(2)} \rangle + \varepsilon^{(1)} | \psi^{(1)} \rangle + \varepsilon^{(2)} | \psi^{(0)} \rangle,
\label{2_order:eqn}
\\
\bigO(\kappa^3): &\!\!\!\!\!\!& \hat H^{(0)} | \psi^{(3)} \rangle + \hat H^{(1)} | \psi^{(2)} \rangle + \hat H^{(2)} | \psi^{(1)} \rangle + \hat H^{(3)} | \psi^{(0)} \rangle \nonumber\\
&\!\!\!\!\!\!&= \varepsilon^{(0)} | \psi^{(3)} \rangle + \varepsilon^{(1)} | \psi^{(2)} \rangle + \varepsilon^{(2)} | \psi^{(1)} \rangle + \varepsilon^{(3)} | \psi^{(0)} \rangle.
\label{3_order:eqn}
\end{eqnarray}

$\hat H^{(0)}$ is the electronic frozen ion Hamiltonian at $X_{\textit{eq}}$ and $\varepsilon^{(0)}$ is the zeroth-order eigenvalue which corresponds to the electronic motion. Therefore, we choose the zeroth-order wavefunction as:
\begin{equation}\label{decouple_two:eqn}
\psi^{(0)}(X_{\textit{eq}},\xi, \zeta) = \chi (\zeta) \varphi^{(0)}(X_{\textit{eq}},\xi),
\end{equation}
where $\varphi^{(0)}$ is the electronic ground state wavefunction of $\hat H^{(0)}$ and $\chi (\zeta)$ is the ionic wavefunction which will be specified later.

Based on Eqs.~(\ref{1_order:eqn}), (\ref{decouple_two:eqn}) and the Hellmann-Feynman theorem, $\varepsilon^{(1)}$ vanishes. This is because the first derivative with respect to the eigenvalue $\varepsilon^{(0)}$ at $X_{\textit{eq}}$ is zero.  More explicitly,  
\begin{equation}\label{first_order_get_zero:eqn}
\begin{split}
\varepsilon^{(1)} &= \langle \psi^{(0)} | \hat H^{(1)} |\psi^{(0)} \rangle \\
&= \left \langle \varphi^{(0)} \left |\left.\frac{\partial}{\partial X} V(X,\xi) \right|_{X=X_{\textit{eq}}} \right |\varphi^{(0)} \right \rangle  \left \langle \chi \left | \zeta \right | \chi
 \right \rangle \\
&= \left.\frac{\partial}{\partial X} \varepsilon^{(0)}(X) \right|_{X=X_{\textit{eq}}} \langle \chi | \zeta | \chi \rangle = 0.
\end{split}
\end{equation}

From Eq.~(\ref{2_order:eqn}) we obtain the second-order correction to the energy:
\begin{equation}
\begin{split}
\varepsilon^{(2)} =& \langle \psi^{(0)}|\hat H^{(2)} |\psi^{(0)} \rangle + \langle \psi^{(0)}|\hat H^{(1)}|\psi^{(1)}\rangle \\
=& \left \langle \chi \left |- \frac{\partial^2}{\partial{\zeta}^2} \right | \chi \right \rangle  \\
&+ \left \langle \varphi^{(0)} \left |\frac{1}{2!} \left.\frac{\partial^2}{\partial X^2} V(X,x,\xi) \right|_{X=X_{\textit{eq}}} \right |\varphi^{(0)} \right \rangle \left \langle \chi \left |\zeta^2 \right |\chi \right \rangle \\
&- \sum_{n>0} \frac{\left | \left \langle \varphi_n^{(0)} \left | \left.\frac{\partial}{\partial X} V(X,x,\xi) \right|_{X=X_{\textit{eq}}} \right | \varphi_0^{(0)} \right \rangle \right |^2} {\varepsilon_n^{(0)} - \varepsilon_0^{(0)}} \langle \chi |\zeta^2| \chi \rangle,
\end{split}
\end{equation}
where the first order correction to the wavefunction is obtained from Eqs.~(\ref{1_order:eqn}) and (\ref{decouple_two:eqn}):
\begin{equation}\label{value_first_order:eqn}
\begin{split}
|\psi^{(1)} \rangle &= - \sum_{n>0} \frac{\left \langle \psi_n^{(0)} \left | \hat H^{(1)} \right | \psi_0^{(0)} \right \rangle } {\varepsilon_n^{(0)} - \varepsilon_0^{(0)}}  | \psi_n^{(0)} \rangle\\
&= - \sum_{n>0} \frac{\left \langle \psi_n^{(0)} \left | \left.\frac{\partial}{\partial X} V(X,x,\xi) \right|_{X=X_{\textit{eq}}} \right | \varphi^{(0)} \right \rangle \zeta | \chi \rangle} {\varepsilon_n^{(0)} - \varepsilon_0^{(0)}}  | \psi_n^{(0)} \rangle.
\end{split}
\end{equation}
Here $\varepsilon_n^{(0)}$ and $| \psi_n^{(0)} \rangle$ are the $n^{th}$ electronic eigenvalue and eigenstate of the Hamiltonian $\hat H^{(0)}$.

We now choose $\chi (\zeta)$ (see Eq.~(\ref{decouple_two:eqn})) to be the lowest eigenfunction of the harmonic oscillator problem. We can then express $\varepsilon^{(2)}$ in the form
\begin{equation}
\varepsilon^{(2)} = \left \langle \chi \left |- \frac{\partial^2}{\partial{\zeta}^2} \right | \chi \right \rangle +
 \frac{1}{2!} k_1  \left \langle \chi \left |\zeta^2 \right | \chi \right \rangle = \frac{1}{2}\omega_I,
\end{equation}
where
\begin{equation}
\begin{split}
k_1 = & \left \langle \varphi^{(0)} \left |\left.\frac{\partial^2}{\partial X^2} V(X,x,\xi) \right|_{X=X_{\textit{eq}}} \right |\varphi^{(0)} \right \rangle \\
& - 2 \sum_{n>0} \frac{\left | \left \langle \varphi_n^{(0)} \left | \left.\frac{\partial}{\partial X} V(X,x,\xi) \right|_{X=X_{\textit{eq}}} \right | \varphi_0^{(0)} \right \rangle \right|^2} {\varepsilon_n^{(0)} - \varepsilon_0^{(0)}},
\end{split}
\end{equation}
is the harmonic oscillator constant. Second order corrections to the energy thus correspond to the ionic vibrations.

Finally, from Eqs.~(\ref{3_order:eqn}), (\ref{decouple_two:eqn}) and (\ref{value_first_order:eqn}) we obtain for the third-order correction to the energy:
\begin{equation}\label{third_vanish:eqn}
\begin{split}
\varepsilon^{(3)} =& \langle \psi^{(0)} | \hat H^{(1)} |\psi^{(2)} \rangle + \langle \psi^{(0)} | \hat H^{(2)} | \psi^{(1)} \rangle + \langle \psi^{(0)} |\hat H^{(3)}|\psi^{(0)} \rangle \\
=& \left \langle \varphi^{(0)} \left |\left.\frac{\partial}{\partial X} V(X,x,\xi) \right|_{X=X_{\textit{eq}}} \right| \varphi^{(2)} \right \rangle  \langle \chi |\zeta^2 | \chi \rangle    
 -\left \langle \chi \left | \frac{\partial^2}{\partial{\zeta}^2} \right | \chi \right \rangle \\
& + 
 \left \langle \varphi^{(0)} \left |\frac{1}{2!} \left.\frac{\partial^2}{\partial X^2} V(X,x,\xi) \right|_{X=X_{\textit{eq}}} \right |\varphi^{(1)} \right \rangle \langle \chi |\zeta^2|\chi \rangle \\
&+ \left \langle \varphi^{(0)} \left |\frac{1}{3!} \left.\frac{\partial^3}{\partial X^3} V(X,x,\xi) \right|_{X=X_{\textit{eq}}} \right | \varphi^{(0)} \right \rangle  \langle \chi |\zeta^3 | \chi \rangle ,
\end{split}
\end{equation}
where the second-order correction to the wavefunction is from Eqs.~(\ref{2_order:eqn}), and (\ref{value_first_order:eqn}):
\begin{equation}\label{second_order_WF:eqn}
\begin{split}
|\psi_0^{(2)}\rangle =& \sum_{n>0} \left (\frac{ \left \langle \psi_n^{(0)} \left | \hat H^{(1)} \right | \psi_0^{(0)} \right \rangle} {\varepsilon_n^{(0)} - \varepsilon_0^{(0)}}\right)^2 | \psi_n^{(0)}  \rangle| \psi_n^{(0)}  \rangle\\
&+ \sum_{n>0} \frac{\left \langle \psi_n^{(0)} \left |\varepsilon^{(2)}-\hat H^{(2)}\right | \psi_0^{(0)} \right \rangle } {\varepsilon_n^{(0)} - \varepsilon_0^{(0)}} | \psi_n^{(0)} \rangle|\psi_0^{(0)} \rangle. 
\end{split}
\end{equation}

All the terms from Eq.~(\ref{third_vanish:eqn}) using Eqs.~(\ref{value_first_order:eqn}) and (\ref{second_order_WF:eqn}) vanish by parity.  This is because they are all proportional to $\langle \psi^{(0)} |\hat H^{(3)} | \psi^{(0)} \rangle$ which is zero by parity since $\hat H^{(3)}$ is odd in $\zeta$ and $ \psi^{(0)}$ is even in $\zeta$.   For this reason $\varepsilon^{(3)} = 0$, and the error in the BOA ground state energy, after including the zero-point energy correction, is $\bigO(\kappa^4) \sim m_e/M$, as shown in Fig.~\ref{FIT_BOA_ERROR:fig}.  

For the neutral homonuclear diatomic molecule we follow the same procedure as above using Eq.~(\ref{quantum_H:eqn}), so that expanded in terms of $\kappa$ gives the Hamiltonian
\begin{equation}\label{H_int:new:eqn}
\begin{split}
\hat H(X_{\textit{eq}} + \kappa \zeta,\xi) \approx &
 - \frac{1}{2\mu_e} \frac{\partial^2}{\partial{\xi}^2}  
+ V(X_{\textit{eq}},\xi)\\
& + \kappa \left.\frac{\partial}{\partial X} V(X,\xi) \right|_{X=X_{\textit{eq}}} \zeta 
 - \kappa^2 \frac{\partial^2}{\partial{\zeta}^2}   \\
& + \frac{1}{2!}\kappa^2 \left.\frac{\partial^2}{\partial X^2} V(X,\xi) \right|_{X=X_{\textit{eq}}} \zeta^2 \\
 &+ \frac{1}{3!}\kappa^3 \left.\frac{\partial^3}{\partial X^3} V(X,\xi) \right|_{X=X_{\textit{eq}}} \zeta^3 
 + \bigO(\kappa^4).
\end{split}
\end{equation}
Again, after including zero-point energy corrections, the error in the BOA ground state energy is $\bigO(\kappa^4) \sim m_e/M$, as shown in Fig.~\ref{FIT_BOA_ERROR:fig}.

\end{document}